\documentclass{article}
\usepackage{amssymb, amsmath, amsthm}
\usepackage{authblk}
\usepackage{a4wide}
\usepackage[bb=dsserif]{mathalpha}
\usepackage{xcolor}
\usepackage{graphicx}
\usepackage{titling}
\usepackage{hyperref}

\graphicspath{{Figures/}}

\newtheorem{theorem}{Theorem}
\newtheorem{lemma}[theorem]{Lemma}
\theoremstyle{remark}
\newtheorem{assumption}{Assumption}
\newcommand{\var}{\textnormal{var}}
\newcommand{\cov}{\textnormal{cov}}
\newcommand{\tr}{\textnormal{tr}}
\newcommand{\ve}[1]{{\bf #1}}
\newcommand{\dd}{\,\text{d}}

\DeclareMathOperator*{\argmin}{arg\,min}

\newcommand{\ex}[1]{\mathbb{E}\left[#1\right]}

\newcommand{\popdt}{\delta t}
\newcommand{\vdt}{\Delta t}
\newcommand{\D}{\mathcal{D}}
\newcommand{\errprob}{\xi}
\newcommand{\timeone}{t}
\newcommand{\timetwo}{t'}
\newcommand{\ddt}{\frac{\partial}{\partial t}}
\DeclareMathOperator*{\trace}{tr}
\newcommand{\swapij}{i \leftrightarrow j}

\title{Consistent diffusion matrix estimation from population time series}
\author{Aden Forrow}
\affil{University of Maine}
\begin{document}
\maketitle
\begin{abstract}
    Progress on modern scientific questions regularly depends on using large-scale datasets to understand complex dynamical systems.
    An especially challenging case that has grown to prominence with advances in single-cell sequencing technologies is learning the behavior of individuals from population snapshots.
    In the absence of individual-level time series, standard stochastic differential equation models are often nonidentifiable because intrinsic diffusion cannot be distinguished from measurement noise.
    Despite the difficulty, accurately recovering diffusion terms is required to answer even basic questions about the system's behavior.
    We show how to combine population-level time series with velocity measurements to build a provably consistent estimator of the diffusion matrix.
\end{abstract}



\section{Introduction}
Learning dynamics from data is a ubiquitous problem in mathematical biology. 
In contexts from cancer diagnoses to ecological forecasts,
we seek to understand how a system will evolve from the current state we observe.
For cell biology, advances in sequencing technologies have led to an explosion of data on the states of individual cells, with concurrent development of new computational algorithms to translate those measurements to predicted dynamics.

Inference algorithms in this field face two key challenges. First, the highest-throughput single-cell sequencing experiments are destructive. 
When cells are lysed to count their RNA content, we lose the possibility of observing what those cells would have done if they had not been measured.
Second, neither the measurements nor the underlying biology is deterministic. 
The data include significant biological variability, which is part of the dynamics we aim to infer, and technical noise from the measurement process, which is not.

These experimental constraints make it impossible to infer time dependence from only a single snapshot of a population without strong assumptions and prior knowledge~\cite{Weinreb2018}.
Recently developed algorithms avoid some of these identifiability problems by considering new types of data, such as population time series~\cite{Schiebinger2019,Wang2022}, RNA velocity~\cite{Bergen2020,LaManno2018,Lange2022,Qiu2022}, and lineage tracing~\cite{Forrow2021,Ventre2023,Wang2022}.
The most important dynamic feature where the state of the art still relies wholly on prior knowledge is the level of biologically meaningful randomness in the system. 
Such randomness can be modeled mathematically by the diffusion matrix $\D$ of a stochastic differential equation (SDE).
This paper provides a provably consistent estimator of $\D$ from population time series measured with velocities.

\subsection{Background}
In the It\^o SDEs we consider, a state $\ve{X}(t) \in \mathbb{R}^d$ evolves according to 
\begin{equation}
    \dd\ve{X}(t) = \ve{v}(\ve{X}(t),t) \dd t + \sigma(\ve{X}(t), t) \dd\ve{B}(t),
    \label{e:SDE}
\end{equation}
where the drift field $\ve{v}(\ve{x},t) \in \mathbb{R}^d$ describes deterministic dynamics and $\sigma(\ve{x}, t)\in \mathbb{R}^{d\times d}$ encodes the stochasticity of the system via the diffusion matrix $\D(\ve{x},t) = \frac{1}{2}\sigma(\ve{x}, t)\sigma(\ve{x}, t)^\top$.
The particle-level SDE can be transformed into a partial differential equation for the evolution of the probability distribution $p(\ve{x}, t)$ of the particles, called a Fokker-Planck equation:
\begin{equation}
    \ddt p(\ve{x}, t) = - \sum_{i=1}^d \frac{\partial}{\partial x_i} \left(v_i(\ve{x},t) p(\ve{x}, t)\right)
    + \sum_{i, j = 1}^d \frac{\partial^2}{\partial x_i \partial x_j}\left( \D_{ij}(\ve{x},t) p(\ve{x},t)\right).
    \label{e:Fokker-Planck}
\end{equation}

We are interested in the inverse problem of determining the terms in the governing equation from experimental observations. 
A broad class of approaches in this area, including SINDy~\cite{Brunton2016} for dynamical systems and SFI~\cite{Frishman2020} for SDEs, applies regression models of various types to time series at the individual level.

When measurements are destructive, however, it is not possible to observe the evolution of one cell's state over time.
Instead, classic RNA sequencing experiments measure collect samples of a population $\{\ve{X}_i(t)\}$.
A time course of $p(\ve{x}, t)$ at the population level can be created by preparing multiple tissues under identical conditions and sequencing each at a different time.
More recent techniques such as RNA velocity~\cite{LaManno2018} and metabolic labeling~\cite{Battich2020,Qiu2022} add noisy measurements of $\ve{v}(\ve{X}(t), t)$, albeit involving extensive modeling~\cite{Bergen2020,Li2020} with important concerns about accuracy~\cite{Gorin2022}.

The challenge of inferring dynamics from large-scale population measurements with limited individual resolution, though particularly prominent for RNA sequencing studies, recurs elsewhere.
In the United States, insurance claims provide a wealth of population-level health information, but tracking individuals with changing insurance arrangements over time both is difficult and raises privacy concerns.
Understanding the behavior of flocking birds~\cite{Cavagna2013} or particles dispersing in the ocean~\cite{Chen2022} is complicated by the difficulty of identifying individuals.

For RNA sequencing, a field of trajectory inference has developed around this inference problem, with methods based on pseudotemporal ordering~\cite{Qiu2017,Street2018}, fitting couplings or Markov chain transition kernels between timepoints~\cite{Forrow2021,Lange2022,Lavenant2024,Schiebinger2019,Wang2022,Weinreb2018}, and kernel regression~\cite{Qiu2022}, among others.
So far, however, no approach has attempted to fit a full SDE including the diffusion term.
Instead, those that quantitatively estimate dynamics  either focus on drift alone~\cite{Qiu2022} or require the user to set a hyperparameter corresponding directly or indirectly to the level of biological stochasticity~\cite{Forrow2021,Lange2022,Lavenant2024,Schiebinger2019,Wang2022,Weinreb2018}.

Some methods skip the diffusion question entirely.
For example, pseudotime approaches such as Monocle~\cite{Qiu2017} or Slingshot~\cite{Street2018} aim to reconstruct the topological structure of a developmental process rather than dynamics in real time.
Dynamo~\cite{Qiu2022} prioritizes learning and analyzing the drift field $\ve{v}(\ve{x},t)$ and avoids making predictions that depend on $\D$, apart from a least action path calculation that assumes $\D$ is constant and isotropic.

Other approaches route around the difficulty of inferring diffusion with mathematical assumptions and tunable parameters.
Weinreb et al.~\cite{Weinreb2018} carefully laid out why a full SDE cannot be identified from a single population snapshot and proposed a method, population balance analysis (PBA), that requires the diffusion matrix as an input parameter.
CoSpar~\cite{Wang2022} makes a foundational assumption that cell fate maps are coherent and sparse; imposing sparsity implicitly constrains the diffusion term not to be too large.

Waddington-OT (WOT)~\cite{Schiebinger2019} and its descendants~\cite{Forrow2021,Lavenant2024,Ventre2023,Zhang2021} rely on entropically regularized optimal transport to couple consecutive timepoints. The regularization parameter corresponds directly to a scalar diffusion coefficient~\cite{Lavenant2024}.
In the absence of direct data on $\D$, WOT uses a computationally convenient but biologically unjustified heuristic based on pairwise distances between cells at distinct timepoints.
Recent mathematical treatments of optimal transport methods~\cite{Lavenant2024,Ventre2023} assume that the diffusion term is known.

CellRank~\cite{Lange2022} has several parameters that govern the degree to which cell state transitions deterministically follow velocities. 
One, $\sigma$, is set using the median Pearson correlation between velocity vectors and state change vectors, which Lange et al. argue helps correct for sparse sampling of velocity vectors.
The mathematical connection between those correlations and the biological stochasticity $\sigma$ ideally would model is unclear. 
Another parameter, $\lambda$, is described as reducing sensitivity to noisy velocity vectors and set to 0.2 by default with no dependence on the data.

These heuristics and assumptions are used not because other authors are unaware of the value of measuring stochasticity directly, but because learning $\D$ from the data available has proven challenging.
Our goal in this paper is to provide an estimator of $\D$ that can fill in the gap in any of the above methods.

\subsection{Paper outline}

Section~\ref{s:theory} presents our theoretical approach. We begin (\S\ref{s:theory_motivation}) with a motivating example where long-term fate probabilities depend dramatically on diffusion. We then review obstacles to identifying diffusion from velocity (\S\ref{s:velocity_nonidentifiability}) or population (\S\ref{s:population_nonidentifiability}) measurements alone. In Section~\ref{s:joint_identifiability}, we derive a consistent estimator of $\ex{\D(\ve{X}, t)}$ from a pair of population measurements with velocities.

In Section~\ref{s:examples} we explore the accuracy of our method in two simulated examples and one application to a dataset of gene expression in the mouse hippocampus.
We conclude with a discussion of future directions in Section~\ref{s:conclusion}.

We use boldface to indicate vectors. Capital letters generally indicate matrices or random variables. 
Thus $\ve{X}$ is a random variable with domain $\mathbb{R}^d$, while $\ve{x}\in \mathbb{R}^d$ is deterministic.
$\ve{\bar X}$ is the mean of $n$ iid samples of $\ve{X}$.


\section{Theory}
\label{s:theory}

\subsection{Relevance of diffusion for fate prediction}
\label{s:theory_motivation}

Our goal is to estimate $\D$. 
Past work~\cite{Bergen2020,Qiu2022} has used RNA velocity~\cite{LaManno2018} and metabolic labeling~\cite{Battich2020} to generate estimates of $\ve{v}(\ve{x},t)$, which, while important, cannot fully answer the biological questions of interest.
With small diffusion, $\ve{X}(t)$ follows a nearly deterministic path to a stationary state or off to infinity; initial conditions therefore determine cell fates.
With large diffusion, fates are entirely unspecified early on.
In between, changing the diffusion parameter can change not just the level of commitment to a fate but a cell's most likely final fate.

\begin{figure}
    \includegraphics[width=0.5\linewidth]{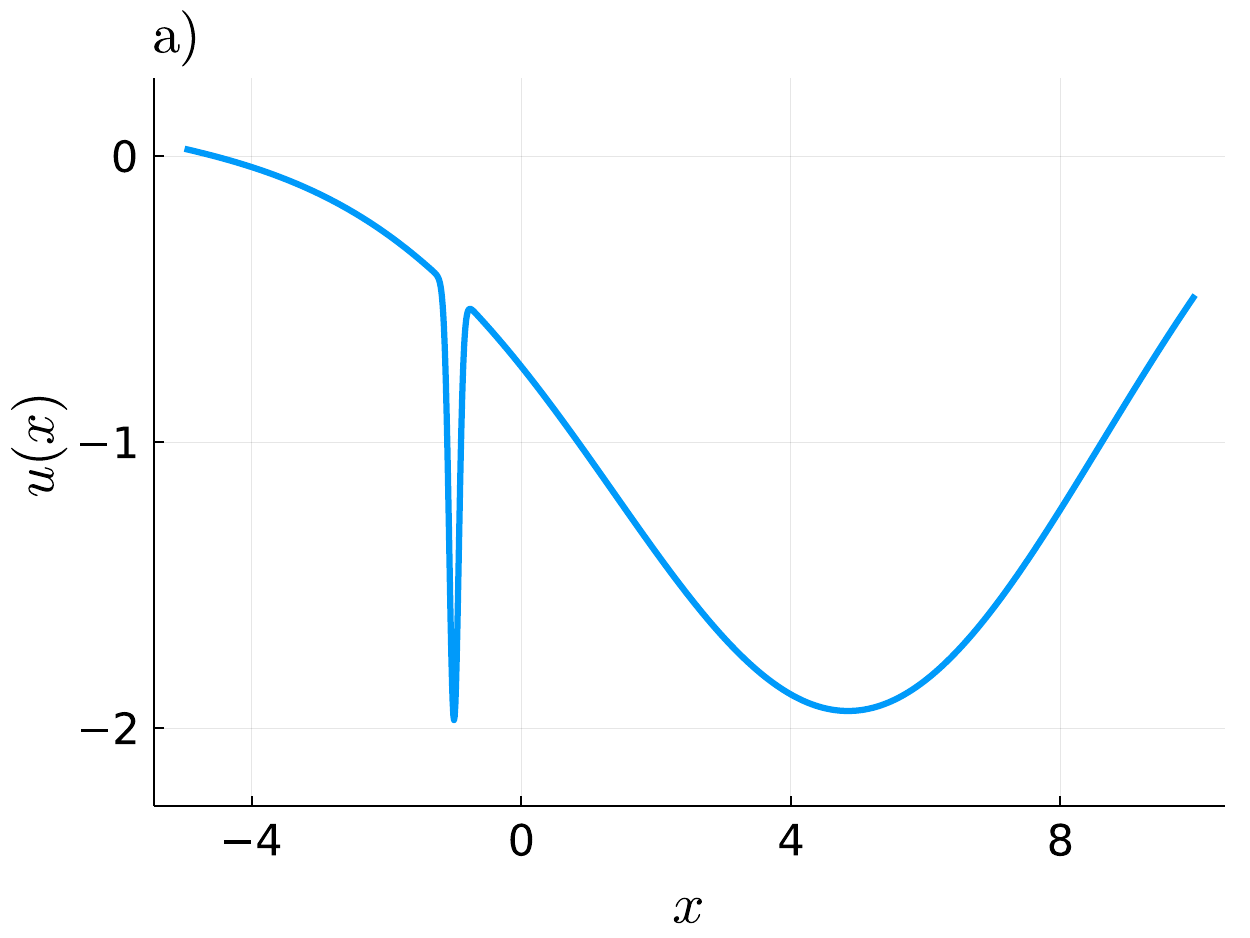}
    \includegraphics[width=0.5\linewidth]{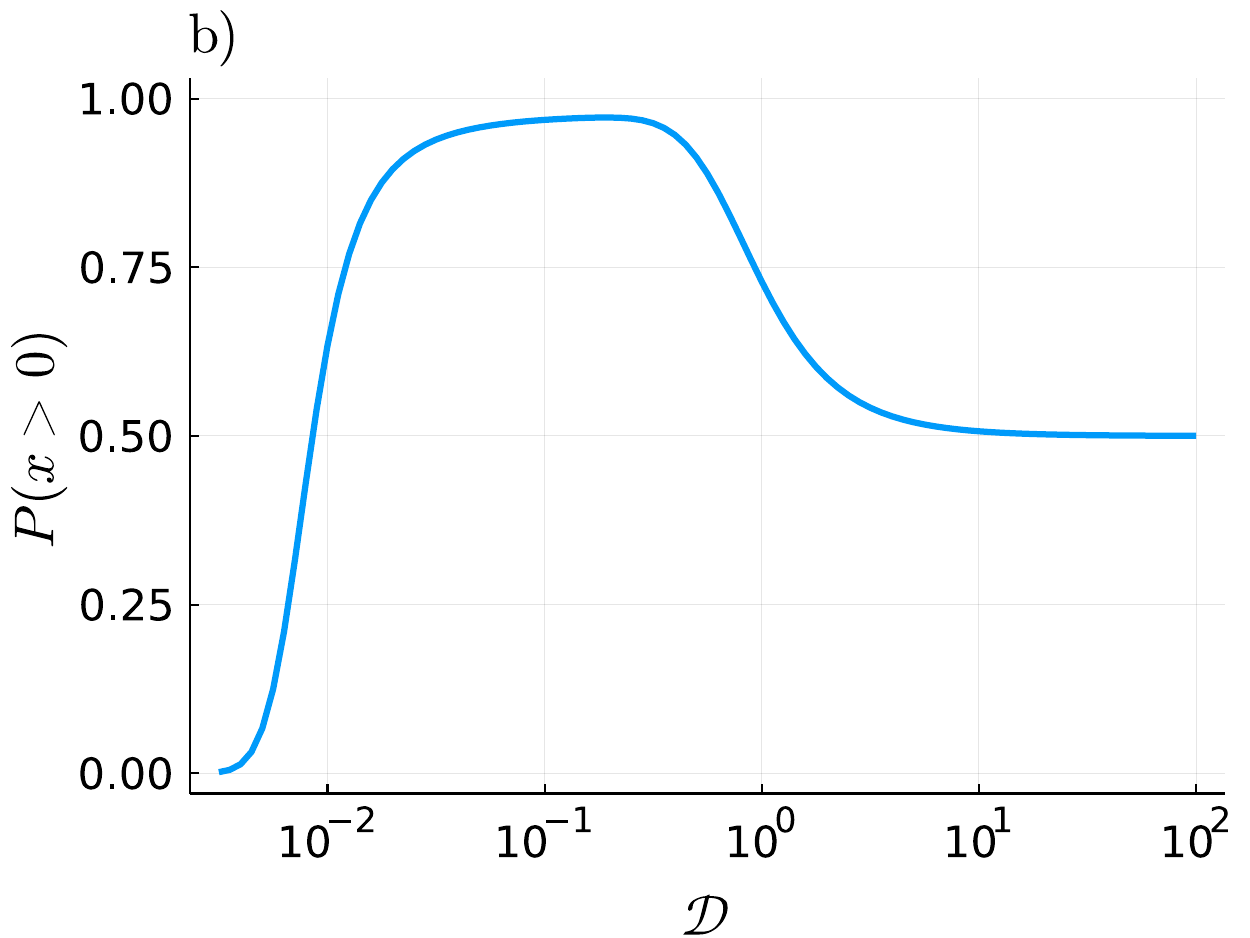}
    \caption{
        Fate probabilities may vary wildly with the diffusion coefficient. 
        a) A one-dimensional potential with two wells, one narrow and deeper and one wide and slightly shallower. 
        b) The probability that $x>0$ in the limiting stationary distribution for the potential in a) ranges from nearly 0 with small $\D$ to over 95\% with moderate $\D$ then drops to 50\% as $\D$ goes to infinity.
    } 
    \label{f:stationary_prob_v_diffusion}
\end{figure}

We illustrate this dependence on $\D$ in Fig.~\ref{f:stationary_prob_v_diffusion} with an example in $d=1$ dimension where we fix a potential 
\begin{align}
    u(x) &= \frac{x^2}{400}
    - \frac{3}{2} \exp\left( - \left(10(x + 1)\right)^2\right)
    - 2 \exp\left( - \left(\frac{x - 5}{5}\right)^2\right),
\end{align}
set $v(x, t) = - u'(x)$, and vary $\D$.
The potential has two wells, one wide at $x\approx 5$ and one narrow and slightly deeper at $x\approx -1$ (Fig.~\ref{f:stationary_prob_v_diffusion}a).
We divide the system into two macrostates, $x> 0$ and $x<0$, and calculate the probability a cell is in each in the stationary distribution $p_\infty(x) \propto \exp(-u(x)/\D)$.

When $\D$ is very small, $p_{\infty}(x)$ is concentrated at the minimum of $u(x)$ in the deeper well, making the probability of $x >0$ near zero (Fig.~\ref{f:stationary_prob_v_diffusion}b).
As $\D$ rises, the difference in depth between the two wells matters less.
Because the shallower well is broader, it grows to dominate the distribution until $P(x > 0) \approx 0.97$.
For still higher $\D$, the symmetric quadratic term dominates and the two states are equally likely.
Clearly, we cannot reasonably claim to know what a cell will do in this system without information about $\D$.
For well-studied systems where fate probabilities are known, they can be used to determine $\D$~\cite{Weinreb2018}; more often, learning the fate probabilities is a key experimental goal.

Despite the importance of $\D$ for answering practical questions about gene expression dynamics, estimating $\D$ is challenging due to statistical identifiability issues with standard experiments. We next review why neither single snapshots with instantaneous velocities nor time courses without velocities have yet permitted statistically consistent estimation of $\D$.

\subsection{Nonidentifiability from velocities}
\label{s:velocity_nonidentifiability}

Several methods~\cite{Bergen2020,Lange2022,Qiu2022} estimate trajectories from measurements of a single population with individual velocities.
The challenge for inferring $\D$ from this data is that biologically meaningful stochasticity may not be distinguishable from technical noise.

In the SDE model, the change  $\Delta\ve{X}(t) = \ve{X}(t + \vdt) - \ve{X}(t)$ in gene expression for a small time interval $\vdt$ is given in terms of the drift, diffusion, and a random increment $\eta \sim \mathcal{N}(\ve{0},I\vdt)$ by
\begin{equation}
\Delta \ve{X}(t) = \ve{v}(\ve{X}(t), t) \vdt + \sigma(\ve{X}(t), t) \ve{\eta} + O(\vdt^{3/2}).
\label{e:finite_SDE_steps}
\end{equation}
Hence RNA velocity and similar techniques, if interpreted as observations of short term increments $\Delta \ve{X}(t)$, do give information about diffusion.

However, each measurement introduces additional noise $\epsilon$.
Rather than $\Delta \ve{X}(t)$, all that can be observed is
\begin{equation}
\Delta \ve{Y} = \Delta \ve{X}  + \ve{\epsilon}.
\end{equation}
Even in the best case where the measurements are unbiased ($\mathbb{E}[\epsilon] = \ve{0}$), the diffusion matrix is not identifiable. If
\begin{equation}
\epsilon \sim \mathcal{N}\left(\ve{0}, \Sigma_{\epsilon}\right),
\end{equation}
then the distribution of observations is 
\begin{equation}
\Delta \ve{Y} \sim \mathcal{N}\left(\ve{v}(\ve{X}(t), t) \vdt, 2\D(\ve{X}(t), t) \vdt +  \Sigma_{\epsilon} \right).
\end{equation}
The mean identifies the drift function up to the time scale $\vdt$, but the covariance matrix identifies only the sum $2\D(\ve{x},t) \vdt +  \Sigma_{\epsilon}$ not the individual terms.
When $\mathbb{E}[\epsilon] = \ve{0}$, methods like Dynamo~\cite{Qiu2022} can consistently estimate $\ve{v}(\ve{x}, t)$ by regressing out both biological and technical noise, but separating the two types to recover $\D$ requires different data or a different model.

\subsection{Nonidentifiability from population time series}
\label{s:population_nonidentifiability}

A parallel series of algorithms, beginning with Waddington-OT~\cite{Schiebinger2019} and continuing with recent work adding deeper mathematical theory~\cite{Lavenant2024,Ventre2023} or incorporating various forms of lineage tracing information~\cite{Forrow2021,Ventre2023,Wang2022}, aim to recover gene expression dynamics from population snapshots at a series of distinct timepoints.
These tools model the data as samples $\ve{X}(t) \sim p(\ve{x}, t)$.
Unfortunately, as is well known in the field~\cite{Weinreb2018}, the population-level marginals $p(\ve{x}, t)$ are insufficient to identify all of the terms in the Fokker-Planck equation, Eq.~\eqref{e:Fokker-Planck}.

The standard nonidentifiability example is a stationary population where $p(\ve{x}, t) = p(\ve{x})$.
As noted by Weinreb et al.~\cite{Weinreb2018}, in a stationary setting replacing $\ve{v}(\ve{x}, t)$ with $\ve{v}(\ve{x}, t) + \ve{\tilde v}(\ve{x}, t)$ where $\nabla_{\ve{x}}\cdot (\ve{\tilde v}(\ve{x}, t) p(\ve{x})) = 0$ gives another SDE with the same stationary distribution.
Similarly, any choice of $\D$ gives a PDE for a corresponding $\ve{v}(\ve{x})$ for which $p(\ve{x})$ is stationary. 

The nonidentifiability of stationary distributions directly implies the existence of nonidentifiable dynamic $p(\ve{x},t)$. If $\ve{x} = (\ve{y}, \ve{z})$ and $p(\ve{x},t) = q(\ve{y},t)r(\ve{z})$ factors into independent parts where one factor is stationary, the components of $\ve{v}$ and $\D$ corresponding to the $\ve{z}$ coordinates could be adjusted similarly to a case where the stationary distribution was $r(\ve{z})$ alone.
A change of coordinates could remove the factorization without changing whether the model is identifiable.

Past work has addressed these identifiability challenges with stronger mathematical assumptions.
Weinreb et al.~\cite{Weinreb2018} use an isotropic $\D$ as an input to PBA and tune it using prior knowledge of fate probabilities.
Lavenant et al.~\cite{Lavenant2024} show that marginals are sufficient to recover the law on paths (and hence the SDE) assuming that (1) $\D = \sigma^2 I$ for known $\sigma^2$ and (2) the flow field is the gradient of a potential.
The requirement that $\D$ be isotropic is not restrictive, as it could be achieved by changing coordinates.
The assumption that $\D$ is known, however, is critical for their consistency results.
The potential assumption rules out some important biological processes, such as the cell cycle.

\subsection{Identifiability from joint measurements}
\label{s:joint_identifiability}

While diffusion is not identifiable from either measurement type individually, combining them allows statistically consistent identification. 
Over a short timescale $\popdt$, state changes for individual cells follow Eq.~\eqref{e:finite_SDE_steps}. Sampling $\ve{X}(t + \popdt)\sim p(\ve{x}, t + \popdt)$ should therefore be approximately equivalent to sampling $\ve{X}(t) \sim p(\ve{x}, t)$, adding $\ve{v}(\ve{X}(t), t)\popdt$, and adding Gaussian noise $\zeta$ with covariance matrix $2\D(\ve{X}(t), t)\popdt$.
We can construct an estimator of $\D(\ve{x}, t)$ by matching these two distributions. 

\begin{figure}
    \includegraphics[width=\linewidth,clip=true,trim={0 0.25in 0 0.4in}]{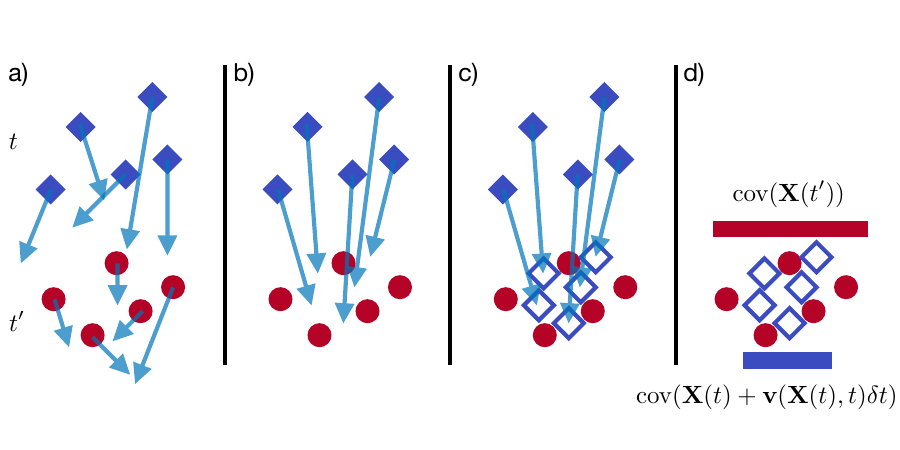}
    \caption{
        Overview of the diffusion estimation process.
        a) We start from observations of $\ve{X}(\timeone)$ (blue diamonds) and $\ve{X}(\timetwo)$ (red circles), together with noisy estimates of the velocities $\ve{v}$ (blue arrows).
        b) We use all the velocity data to make smoothed regression estimates of $\ve{v}(\ve{X}(\timeone), \timeone)$.
        c) Next, we push the observations $\ve{X}(\timeone)$ forward to $\ve{X}(\timeone) + \ve{v}(\ve{X}(\timeone), \timeone)$.
        d) Because pushing forward with only $\ve{v}$ omits the contribution of diffusion, we expect the covariance of the pushforward samples to be smaller than the covariance of the data observed at $\timetwo$. The difference can be used to estimate $\D$.
    }
    \label{f:cartoon}
\end{figure}

For simplicity, in this paper we only estimate a constant diffusion matrix $\D$, although in principle dependence on $\ve{x}$ could be necessary to fit the data. The covariance matrix of the observations at $\timetwo = \timeone + \popdt$ should approximately equal the covariance matrix of the observations at $\timeone$ pushed forward with $\ve{v}(\ve{X}(\timeone), \timeone)$ and $\D(\ve{X}(\timeone), \timeone)$:
\begin{align}
    \cov(\ve{X}(\timetwo)) \approx \cov(\ve{X}(\timeone) + \popdt \ve{ v}(\ve{X}(\timeone), \timeone) + \zeta).
\end{align}
If $\D$ is constant, $\zeta$ is independent of $\ve{X}(t)$ and the right hand side can be expanded to
\begin{align}
    \cov(\ve{X}(\timeone) + \popdt \ve{ v}(\ve{X}(\timeone), \timeone) + \zeta) = \cov(\ve{X}(\timeone) + \popdt \ve{ v}(\ve{X}(\timeone), \timeone)) + 2\D\popdt.
\end{align}
Solving for $\D$ defines an estimator
\begin{align}
    \widehat{\D} & = \frac{1}{2\popdt}\left(\widehat\cov(\ve{X}(\timetwo)) - \widehat\cov(\ve{X}(\timeone) + \popdt \ve{ \hat v}(\ve{X}(\timeone), \timeone))\right),
    \label{e:covariance_difference_D_estimate}
\end{align}
where $\ve{ \hat v}(\ve{x})$ is a regression estimate of $\ve{v}(\ve{x})$ and $\widehat{\cov}(\ve{X})$ denotes the standard unbiased covariance estimator
\begin{align}
    \widehat{\cov}(\ve{X}) & = \frac{1}{n-1}\sum_{i=1}^n (\ve{X}_i - \ve{\bar X}) (\ve{X}_i - \ve{\bar X})^\top.
\end{align}
Figure~\ref{f:cartoon} shows a sketch of how $\widehat\D$ is built.

When $\D(\ve{x}, t)$ may depend on $\ve{x}$ and $t$, $\widehat{\D}$ is a provably consistent estimator of $\ex{\D(\ve{X}(t), t)}$, as we formalize in our core theorem.

\begin{theorem}
    Suppose $p(\ve{x}, t)$ is a solution to Eq.~\eqref{e:Fokker-Planck} such that $\ve{v}(\ve{X}(t), t)$, $\D(\ve{X}(t), t)$, their derivatives to second order in $\ve{x}$ and first order in $t$, and $\ve{X}(t)$ itself have bounded fourth moments with $\ve{X}(t)\sim p(\ve{x}, t)$.
    
    Given $n\ge 2$ iid samples from $p(\ve{x}, t)$ and $n$ iid samples from $p(\ve{x}, t+ \popdt)$,
    \begin{align}
        \|\widehat{\D} - \ex{\D(\ve{X}(t), t)} \|_\infty \le 
         O\bigg(\|\ve{\hat v}(\ve{x}, t) - \ve{v}(\ve{x}, t)\|_\infty + \popdt\bigg) + O_p\left(\frac{1}{\popdt\sqrt{n}}\right).
        \label{e:D_error_bound}
    \end{align}
    \label{thm:consistency}
\end{theorem}
We include the proof of Theorem~\ref{thm:consistency} in the supplement (\ref{s:convergence_proof}).

The three terms on the right of Eq.~\eqref{e:D_error_bound} correspond to three distinct sources of error. The first is due to inaccuracy in $\ve{\hat v}(\ve{x}, t)$; the second to the finite time resolution $\popdt$; and the third to estimating population covariances from a finite number of samples. Importantly, as long as $\ve{\hat v}(\ve{x}, t)$ is a consistent estimator of $\ve{v}(\ve{x}, t)$, all three go to zero in the joint limit $\popdt\to 0$ and $\popdt\sqrt{n} \to \infty$.

Note that there is no guarantee that $\widehat{\D}$ from Eq.~\eqref{e:covariance_difference_D_estimate} is positive definite. In fact, it will never be positive definite when the timestep $\popdt$ is too large. 
Negative eigenvalues of $\widehat{\D}$ are a signal that the estimate is unreliable and better data or modeling is required.

A final consideration for RNA velocity in particular is that the timescale of the velocities may not be known. Counts of spliced and unspliced RNA at time $\timeone$ contain no information about the overall speed of gene expression dynamics, and hence can only provide an estimate $\ve{\hat u} \approx \alpha^{-1} \ve{v}(\ve{x}, \timeone)$ for some unknown $\alpha$. When the process is not stationary, population measurements make $\alpha$ identifiable. Applying the same logic that led to the estimator $\widehat{\D}$ to means instead of covariances,  we expect
\begin{align}
    \ex{\ve{X}(\timetwo)} \approx \ex{\ve{X}(\timeone) + \popdt \alpha \ve{\hat u}}.
    \label{e:mean_change}
\end{align}
Eq.~\eqref{e:mean_change} suggests a least-squares estimator
\begin{align}
    \hat \alpha = \argmin_{\alpha} \| \ex{\ve{X}(\timetwo)} - \ex{\ve{X}(\timeone) + \popdt \alpha \ve{\hat u}} \|^2_2.
    \label{e:velocity_timescale_estimator}
\end{align}
We use $\hat\alpha$ to set the velocity timescale in our examples in Section~\ref{s:examples}.
For stationary populations where $\ex{\ve{\hat u}} = \ve{0}$, velocity data with time information, such as metabolic labeling, may be necessary to determine the timescale.
Moreover, just as $\widehat{D}$ need not be positive definite, $\hat \alpha$ need not be positive.
Observing $\hat \alpha < 0$ suggests $\ve{\hat u}$ is unreliable as it points on average in the opposite direction of the change in the population.


\section{Examples}
\label{s:examples}

We present three examples to illustrate the behavior of $\widehat{\D}$, in each case comparing to the default estimator from WOT. 
First, we validate the dependence of error on $n$ and $\popdt$ in simulations with a linear $\ve{v}(\ve{x},t)$.
Second, we use a simulated two-state system to investigate the dependence of state transition probabilities on $\D$.
Finally, we estimate $\D$ for a dataset of hippocampus gene expression. Although no ground truth is available, our results suggest the WOT default is much too high.

\subsection{$\widehat{\D}$ accuracy with linear drift}

\begin{figure}
    \includegraphics[width=0.5\linewidth]{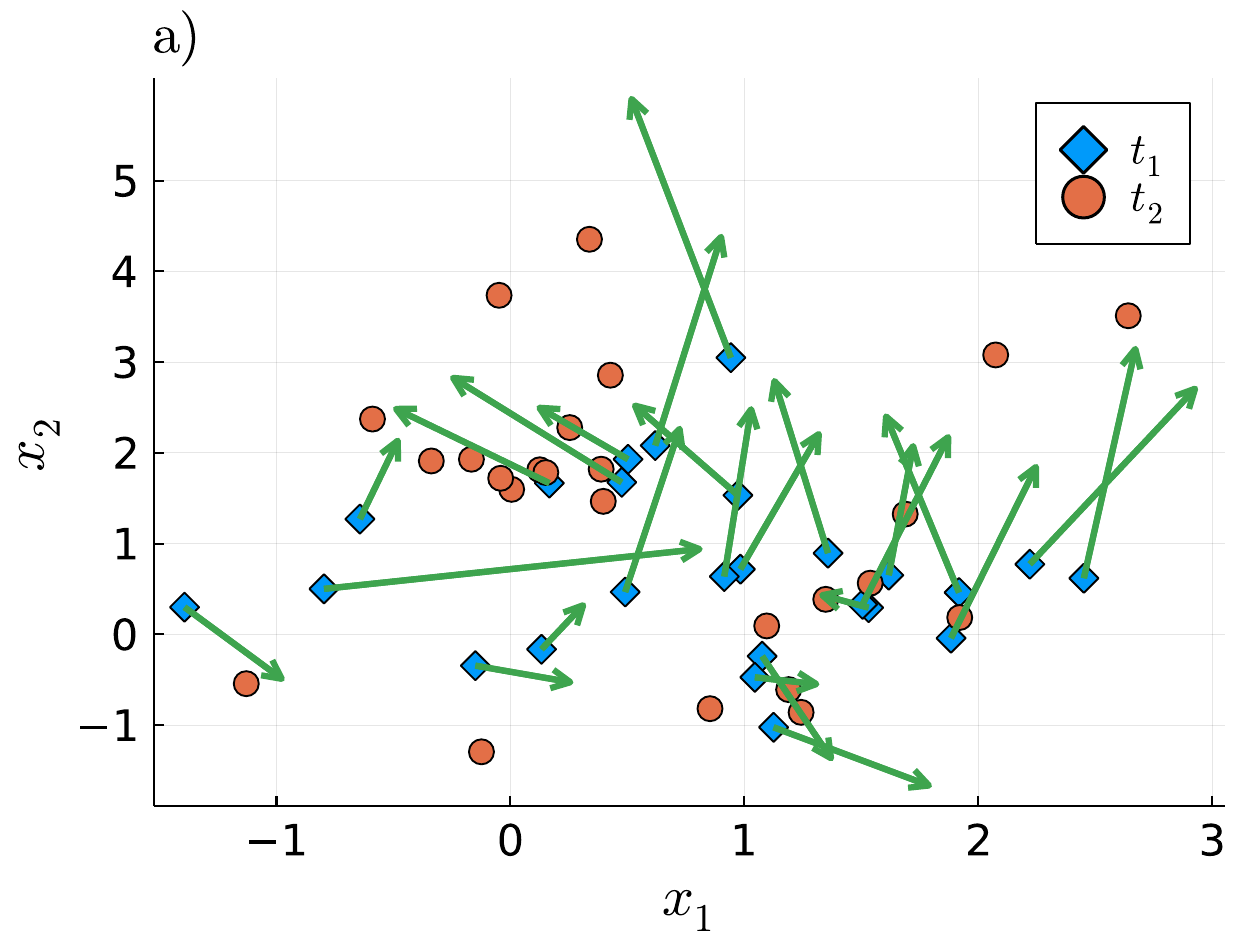}
    \includegraphics[width=0.5\linewidth]{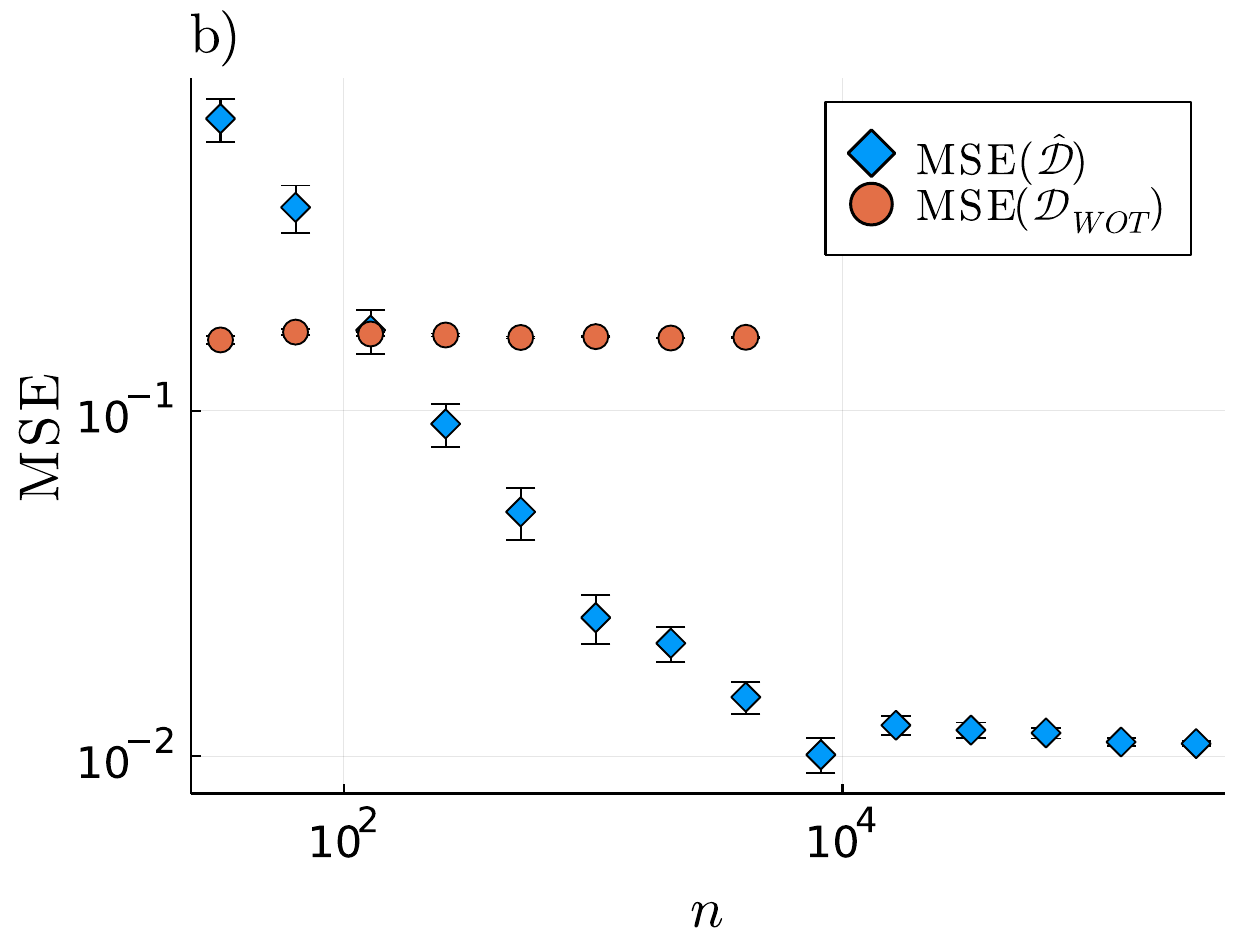}

    \includegraphics[width=0.5\linewidth]{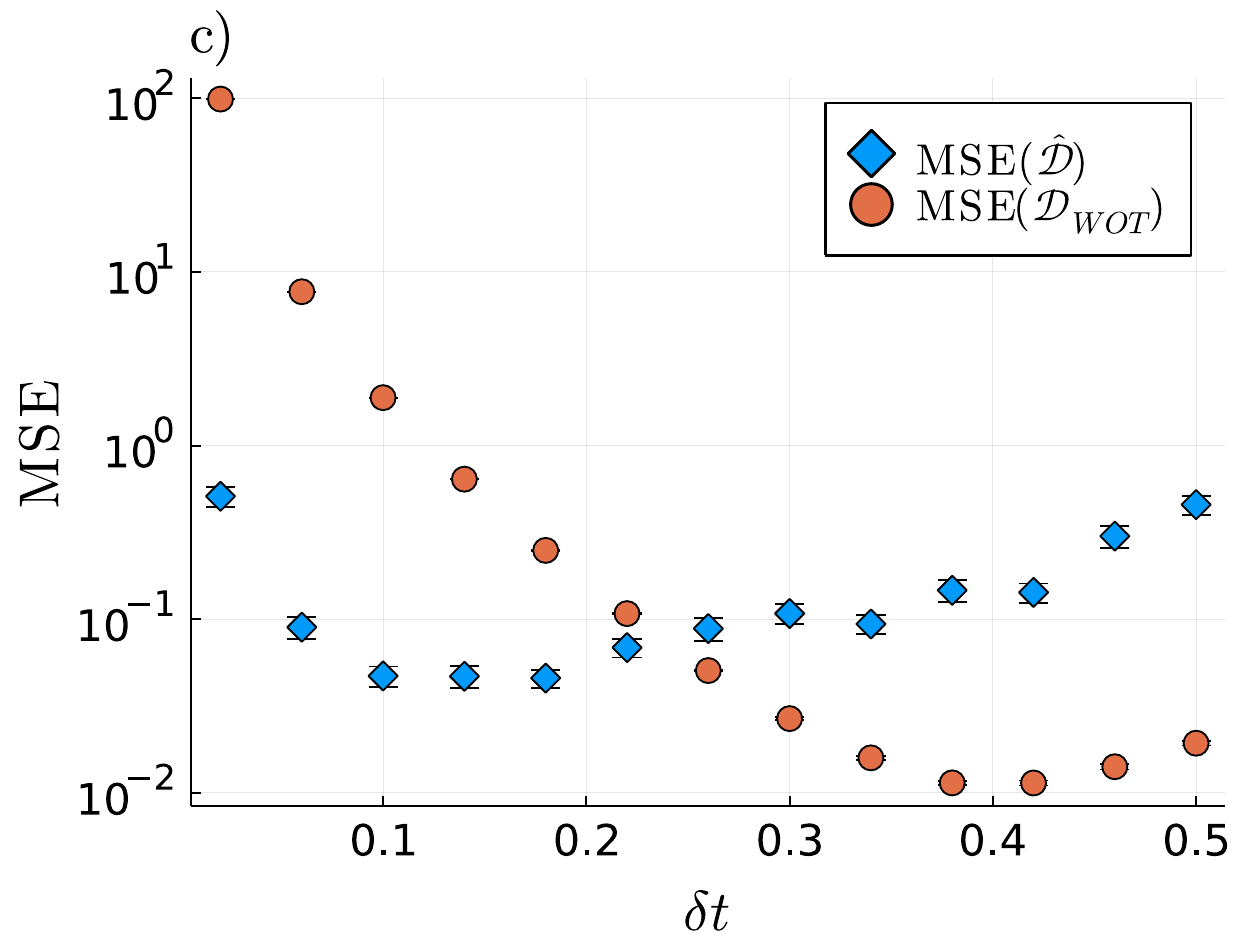}
    \includegraphics[width=0.5\linewidth]{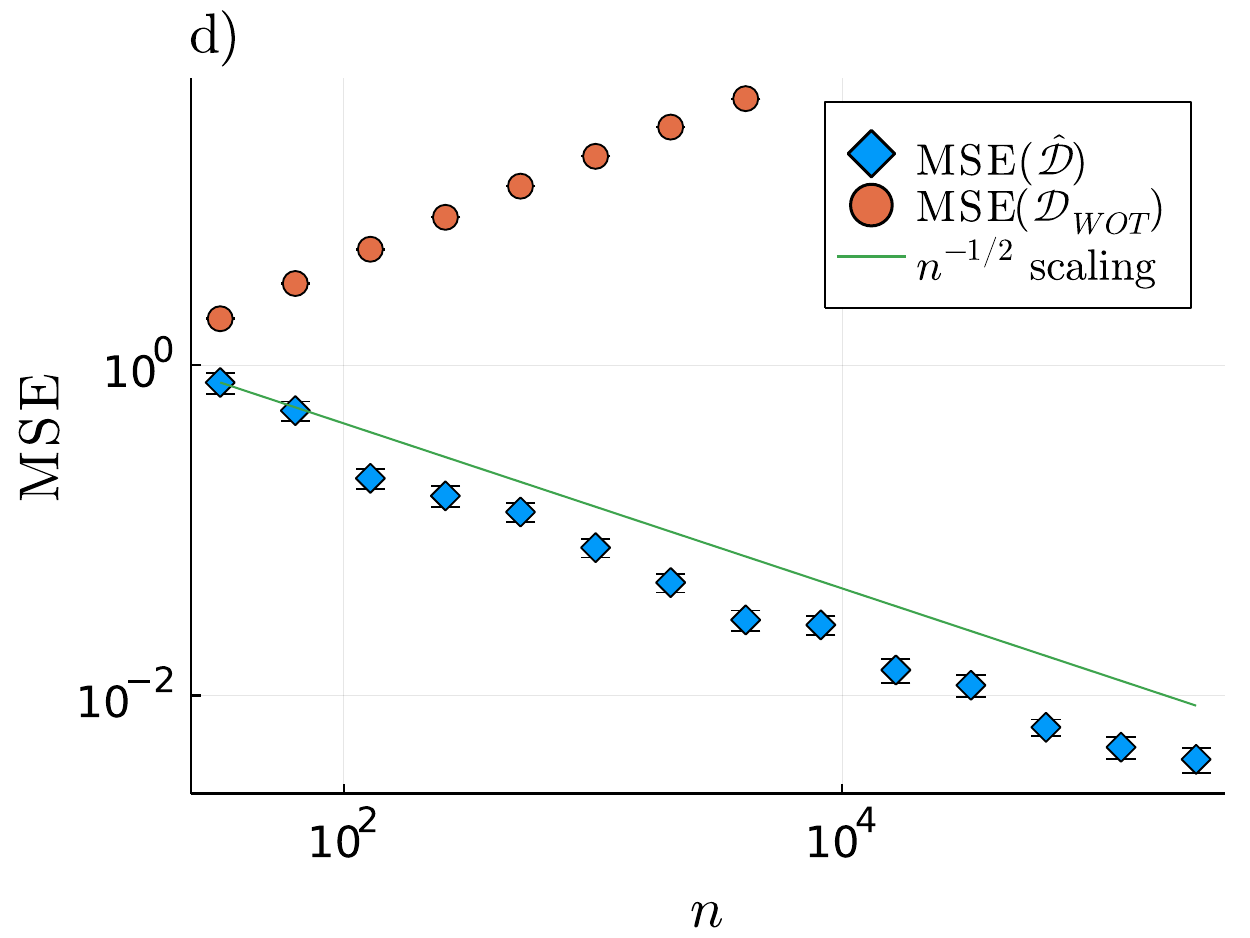}
    \caption{
        The dependence on $\delta t$ and $n$ in Eq.~\eqref{e:D_error_bound} is visible in tests in $d=10$ dimensions with $\mathcal{D} = I$ and $\ve{v}(\ve{x}, t) = A\ve{x}$ where $A$ is a random matrix with iid $\mathcal{N}(0,1)$ entries. The initial $X_i(t)$ are iid $\mathcal{N}(1, 1)$.
        a) The data we use are observations of states and velocities at $\timeone$ as well as states at $\timetwo$, here with $\popdt = 0.2$.
    b) Increasing sample size with fixed $\popdt = 0.2$ reduces the error of $\widehat{\D}$ but does not improve $\D_{WOT}$. 
    As $\D_{WOT}$ is a scalar, we measure accuracy with the MSE of $\textrm{trace}(\D)/d$.
    c) With a fixed sample size $n = 512$, error increases with both too little time $\popdt$ between samples and too much. 
    d) By reducing $\popdt$ as sample size increases (here $\popdt = 0.2 (n/2)^{-1/4}$) $\widehat{\D}$ is asymptotically consistent, removing the plateau in (b).  
    $\D_{WOT}$, on the other hands, diverges as $\popdt \to 0$.
    Points in (b-d) are averages of 100 trials with the same $A$; error bars show $\pm 1$ sample standard errors.
    Because $\D_{WOT}$ does not perform better with larger sample sizes, we omit points for larger values of $n$ to avoid the $O(n^2 d)$ computational cost.
    }
    \label{f:linear_drift_example}
\end{figure}

Our first example, presented in Fig.~\ref{f:linear_drift_example}, has a drift field $\ve{v}(\ve{x}, t) = -A\ve{x}$ in $d = 10$ dimensions with $\mathcal{D} = I$ and initial $\ve{X}(\timeone) \sim \mathcal{N}(\mu(\timeone), \Sigma(\timeone))$, where $\mu(\timeone) = \ve{1}$ is the vector of all ones and $\Sigma(\timeone) = I$.
Entries of the matrix $A\in \mathbb{R}^{10\times10}$ are iid $\mathcal{N}(0,1)$.
If $A$ were symmetric, this would correspond to $\ve{v}(\ve{x}, t) = -\nabla u(\ve{x})$ with potential $u(\ve{x}) = \ve{x}^\top A \ve{x}/2$.

With linear drift and constant diffusion, the SDE Eq.~\eqref{e:SDE} is analytically solvable~\cite{Gardiner2004}. The distribution $p(\ve{x}, \timetwo)$ at the later timepoint $\timetwo$ is also normal with mean
\begin{align}
    \mu(\timetwo) & = e^{-A\popdt} \mu(t)
\end{align}
and covariance matrix satisfying
\begin{align}
    A\Sigma(\timetwo) + \Sigma(\timetwo)A^\top= A e^{-A\popdt}\Sigma(\timeone) e^{-A^\top \popdt} + e^{-A\popdt}\Sigma(\timeone) e^{-A^\top \popdt}A^\top + 2\D - 2e^{-A\popdt}\D e^{-A^\top \popdt}.
    \label{e:linear_drift_covariance_equation}
\end{align}
Eq.~\eqref{e:linear_drift_covariance_equation} is a linear system that can be solved for $\Sigma(\timetwo)$, allowing us to sample from $p(\ve{x}, \timetwo)$ without numerical integration of the SDE.

Figure~\ref{f:linear_drift_example}a shows an example of the simulated data projected down to the first two coordinates. 
To focus on the dependence on $n$ and $\delta t$ we use exact velocities $\ve{\hat{u}}(\ve{x}, t) = \ve{v}(\ve{x},t)$, which are then scaled by $\hat \alpha$ from Eq.~\eqref{e:velocity_timescale_estimator}.

We compare $\widehat{\D}$ to the estimator of $\D$ implicit in the default entropic regularization in WOT.
Mathematically, applying optimal transport with entropic regularization parameter $\epsilon$ corresponds to a diffusion matrix $\D = \epsilon I /\popdt$~\cite{Lavenant2024}.
WOT by default sets $\epsilon$ equal to $0.05$ times the median of the pairwise squared Euclidean distances between $\ve{X}(\timeone)$ and $\ve{X}(\timetwo)$. We write $\D_{WOT}$ for this default $\epsilon$ divided by $\delta t$.
Because WOT, like every other dynamic inference method we are aware of, does not attempt to estimate a diffusion matrix, we only evaluate the accuracy of the scalar $\tr(\D)/d$.

In Figure~\ref{f:linear_drift_example}b-d we present the mean squared error (MSE) of the estimators $\widehat{\D}$ and $\D_{WOT}$ for varying $n$ and $\delta t$.
We evaluate MSE as the empirical average of $100$ trials estimating $\tr(\D)/d$, i.e.
\begin{align}
    \textrm{MSE}(\widehat \D) = \frac{1}{100} \sum_{k=1}^{100} \left(\frac{\tr(\widehat{\D}_k) - \tr(\D)}{d}\right)^2.
\end{align}

With fixed $\popdt$, increasing $n$ reduces the sampling error of $\widehat{D}$ until it is small relative to $O(\popdt)$ bias (Fig.~\ref{f:linear_drift_example}b).
On the other hand, $\D_{WOT}$'s MSE is dominated by bias and does not improve with increasing $n$.
With fixed $n$ (Fig.~\ref{f:linear_drift_example}c), the MSE of $\widehat{\D}$ increases with both too small $\delta t$ (due to the $O(1/\delta t)$ sampling error term in Eq.~\eqref{e:D_error_bound}) and too large $\delta t$ (due to the $O(\delta t)$ bias term in Eq.~\eqref{e:D_error_bound}).
$\D_{WOT}$ can have low error when $\popdt$ is tuned to relate pairwise distances to $\D$; otherwise, it incorrectly diverges as $\popdt$ goes to zero.

Unlike $\D_{WOT}$, $\widehat{\D}$ can be made consistent by taking the joint limit $n \to \infty$, $\popdt \to 0$.
If $\popdt \sim n^{-1/4}$, both the finite-time bias and sampling error in Eq.~\eqref{e:D_error_bound} are $O(n^{-1/4})$.
After squaring, we expect MSE to scale like $n^{-1/2}$ in this limit, which we see in Fig.~\ref{f:dentate_gyrus}d.
Altogether, these results suggest the bound in Thm.~\ref{thm:consistency} accurately captures typical limiting behavior of $\widehat{\D}$.

\subsection{Probabilities of state transitions}

\begin{figure}\includegraphics[width=0.5\linewidth]{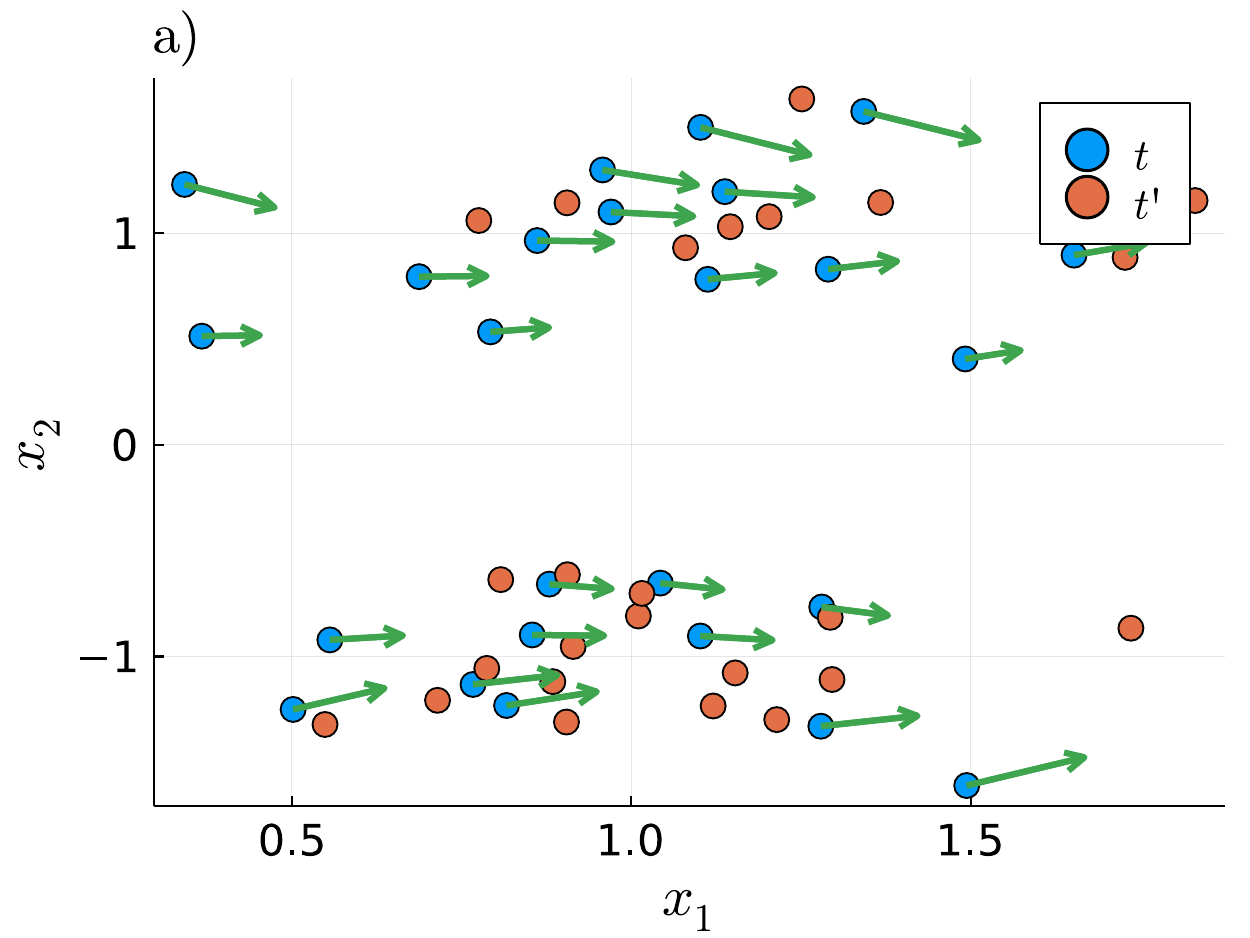}
    \includegraphics[width=0.5\linewidth]{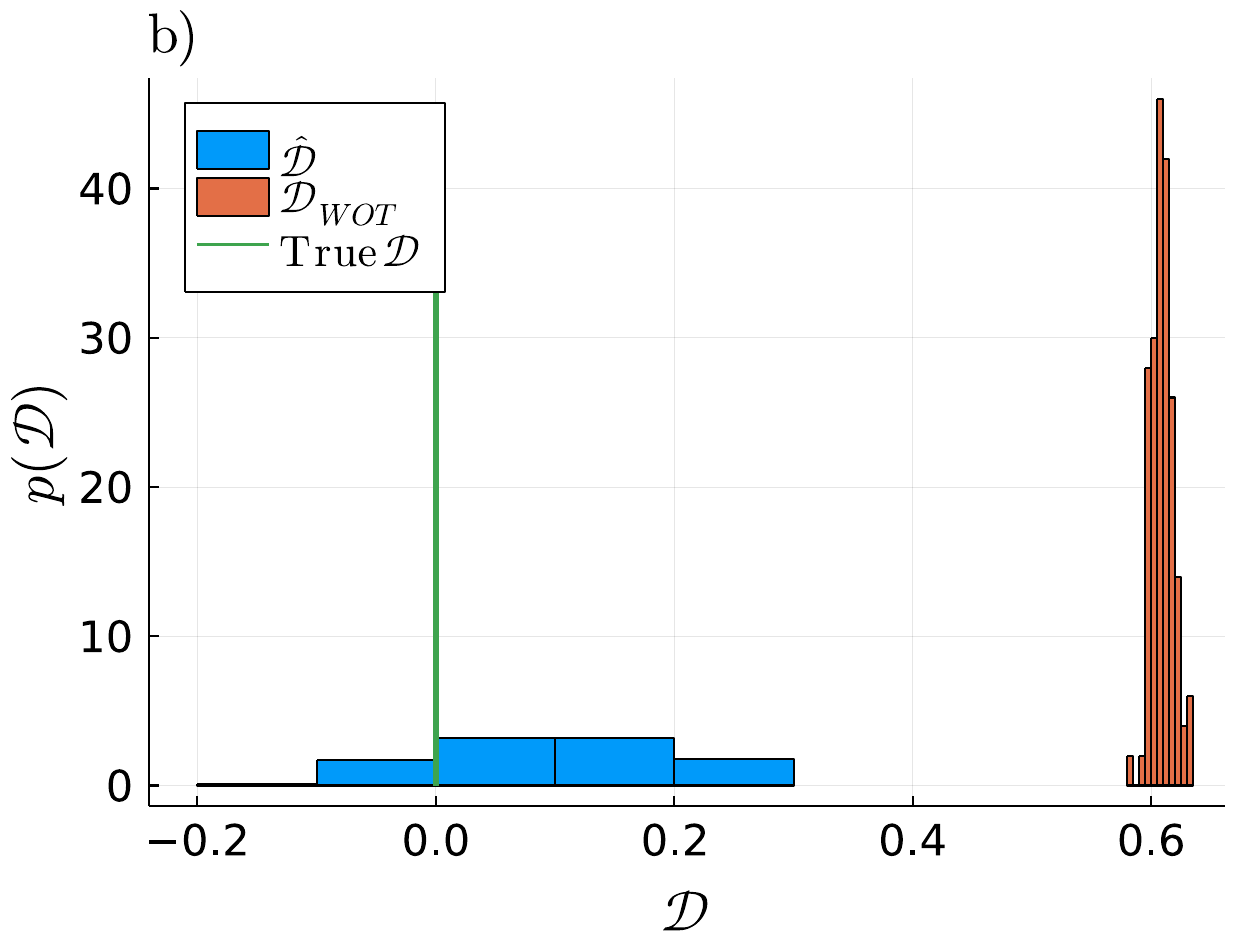}

    \includegraphics[width=0.5\linewidth]{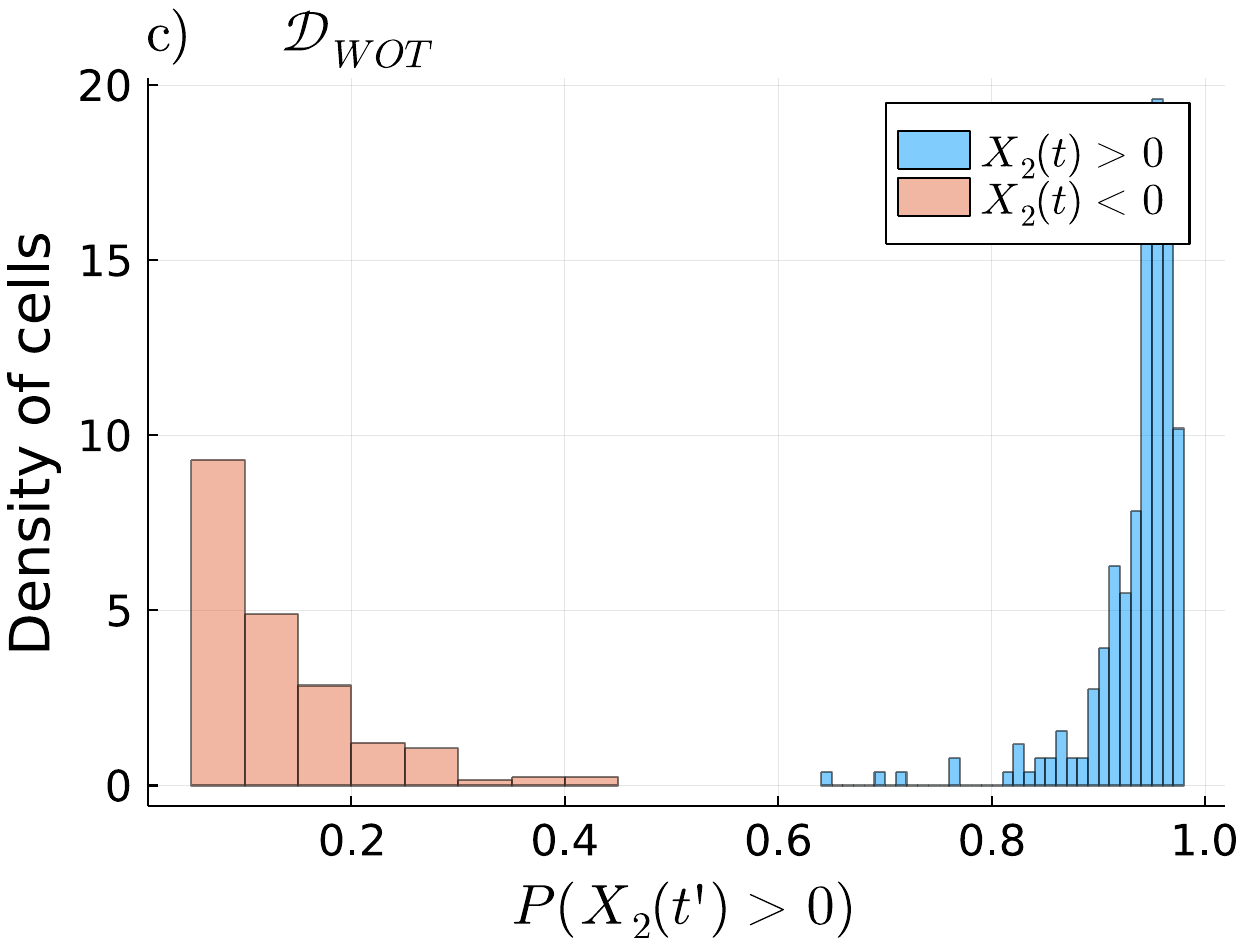}
    \includegraphics[width=0.5\linewidth]{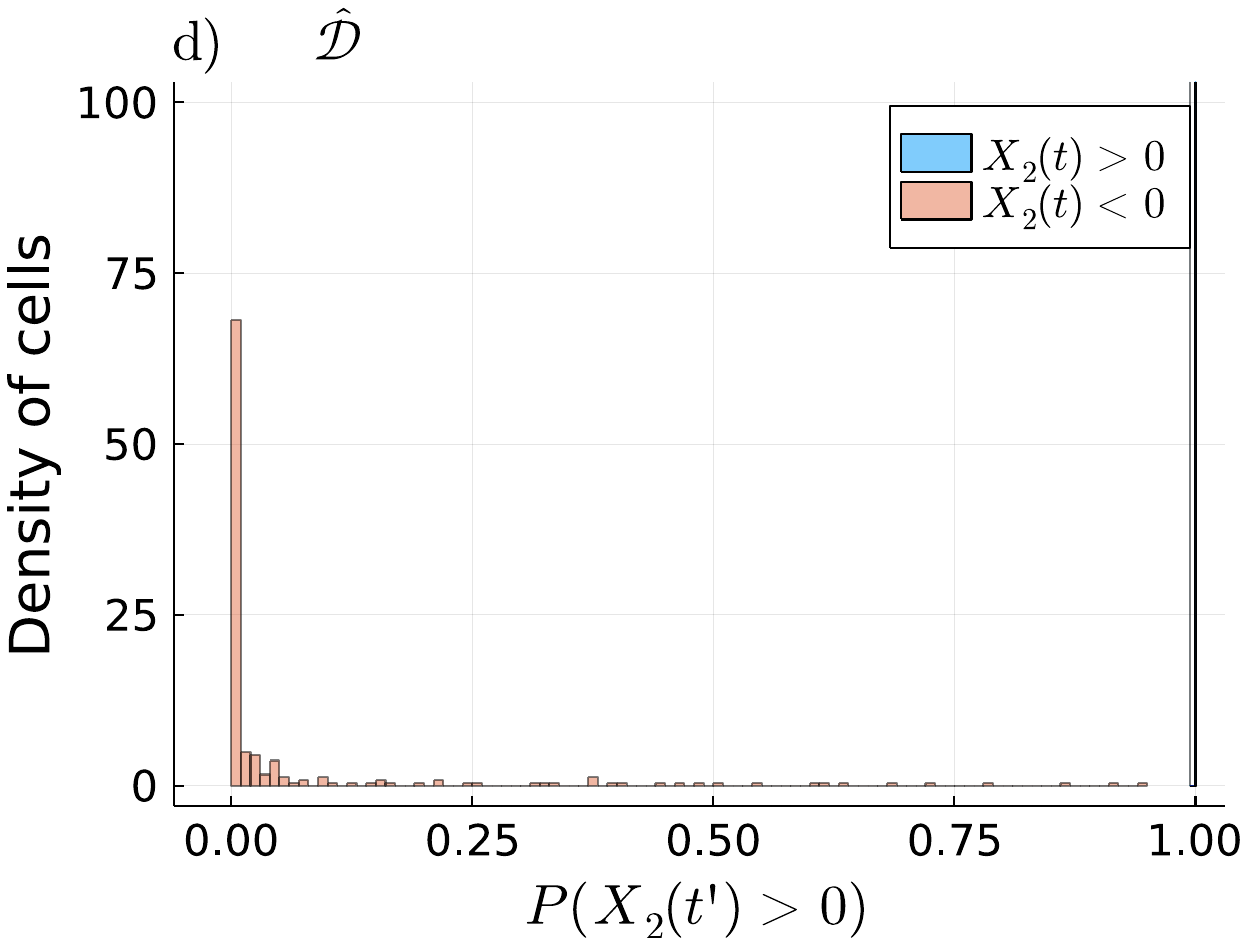}
    \caption{
        Error in estimating $\D$ leads to error in estimating transition probabilities between states.
        a) We simulated a deterministic ($\D = 0$) flow in the two-channel potential of Eq.~\eqref{e:bifurcation_potential}. 
        At $\timeone$, $\ve{X}(\timeone)\in \mathbb{R}^2$ is sampled from an equally weighted mixture of $\mathcal{N}((1,1)^\top, 0.1I)$ and $\mathcal{N}((1,-1)^\top, 0.1I)$. 
        The $\ve{X}(\timetwo)$ samples are initialized from the same distribution and follow $\ve{v}(\ve{x}) = -\nabla u(\ve{x})$ for $\popdt = 0.1$.
        b) The distribution of $\tr(\widehat{\D})/d$ from 100 trials of $n = 500$ samples each is centered slightly above 0, with around a 20\% chance $\tr(\widehat{\D}) < 0$. $\D_{WOT}$ has much lower variance and greater bias.
        c) Because $\D_{WOT}$ is too large, using it to set the entropic regularization parameter for optimal transport incorrectly suggests that many cells transition between the $x_2 > 0$ and $x_2 < 0$ channels in the short interval $\delta t$.
        Because $\D = 0$, no transitions occur in the ground-truth simulations. The correct fate probabilities are $P(X_2(\timetwo) > 0 | X_2(\timeone) > 0) = 1$ and $P(X_2(\timetwo) > 0 | X_2(\timeone) < 0) = 0$.
        d) In a typical example with $\tr(\widehat{\D})/d = 0.09$, the estimated transition probabilities are much more accurate.
    }
    \label{f:bifurcation}
\end{figure}

One of the key dynamic features of interest in time course expression studies is the probability a cell will transition from one state to another.
We explore the relevance of diffusion for this question in a simulation in $d = 2$ dimensions with two parallel channels.
The drift field in Fig.~\ref{f:bifurcation} is $\ve{v}(\ve{x}, t) = -\nabla u(\ve{x})$ with
\begin{align}
    u(\ve{x}) = -x_1 - \frac{1}{2} x_1 x_2^2  + \frac{1}{4} x_2^4.
    \label{e:bifurcation_potential}
\end{align}
In the $x_2$ direction, this potential has two minima at $x_2 = \pm \sqrt{x_1}$.

We draw $x_1(\timeone)$ from $\mathcal{N}(1, 0.1)$ and $x_2(\timeone)$ from an equally weighted mixture distribution of $\mathcal{N}(1, 0.1)$ and $\mathcal{N}(-1, 0.1)$.
The samples are therefore approximately equally split between the states with $x_2 > 0$ and $x_2 < 0$.
We set $\D = 0$, which means transitions between the states never occur.
The later samples $\ve{X}(\timetwo)$ are generated by sampling $\ve{X}(\timeone)$ in the same way and then numerically integrating Eq.~\eqref{e:SDE} for $\delta t = 0.1$.

Figure~\ref{f:bifurcation}a shows an example of the simulated data. The estimates $\widehat{\D}$ with $n = 500$ samples from each timepoint have much lower bias and higher variance than $\D_{WOT}$ (Fig.~\ref{f:bifurcation}b).
In Fig.~\ref{f:bifurcation}c and d we compare transition probabilities between the $x_2 > 0$ and $x_2 < 0$ macrostates if we couple $\timeone$ and $\timetwo$ with entropically regularized optimal transport. 

In the ground truth simulations, no transitions occur because $\D = 0$.
Using $\D_{WOT}$ leads to moderate transition probabilities (Fig.~\ref{f:bifurcation}c).
Over the course of a longer time series, these transitions probabilities would compound; a WOT analysis would therefore incorrectly conclude that fates are not well specified early.
The transition probabilities using $\epsilon = \tr(\widehat{\D})\popdt/d$ (Fig.~\ref{f:bifurcation}d) are much more accurate.
Perfect accuracy is not attainable because we enforce strict marginal constraints when fitting optimal transport couplings and the proportions of cells in each state at $\timeone$ and $\timetwo$ differ due to sampling variation.

\subsection{Hippocampus gene expression}

\begin{figure}
    \includegraphics[width=0.5\linewidth]{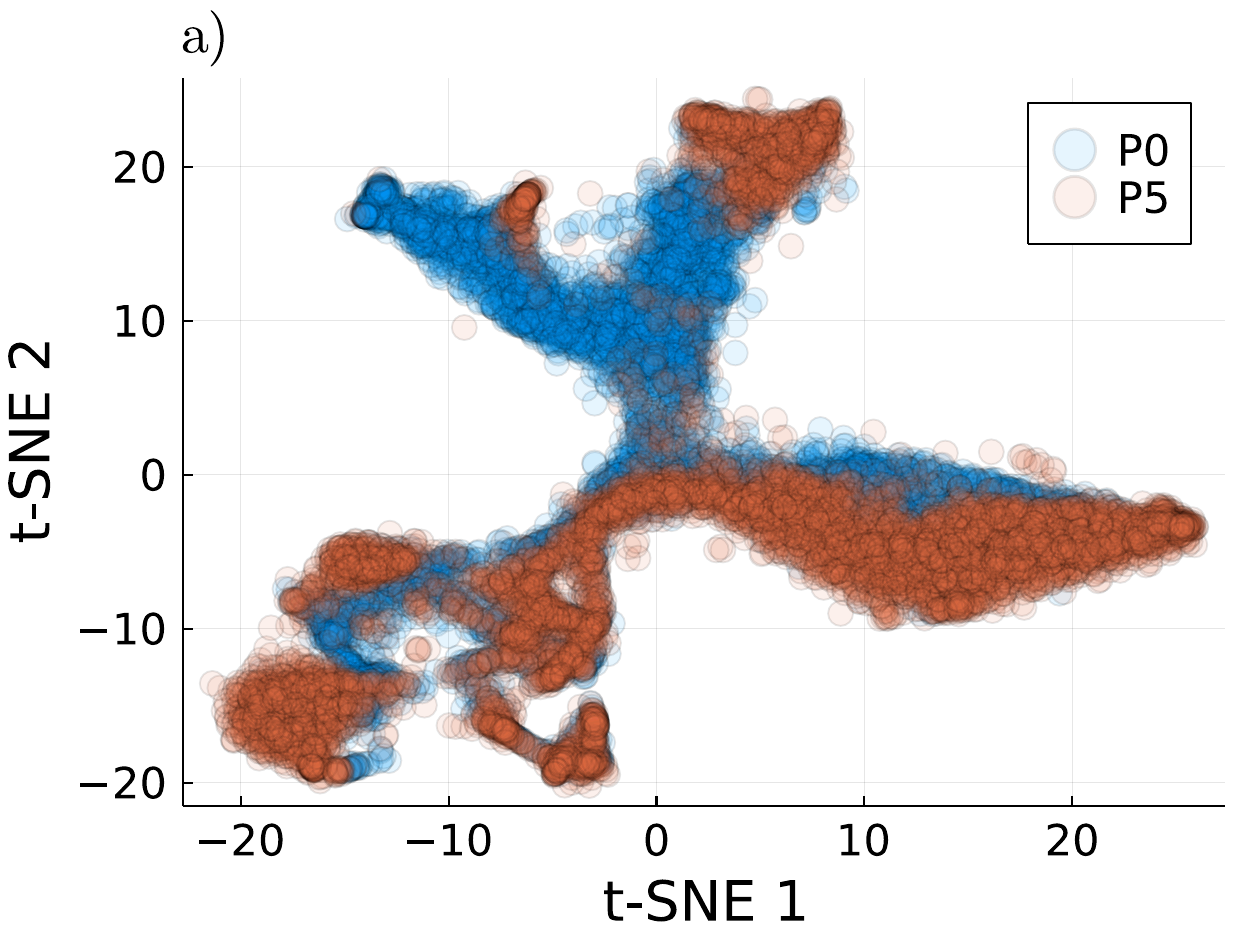}
    \includegraphics[width=0.5\linewidth]{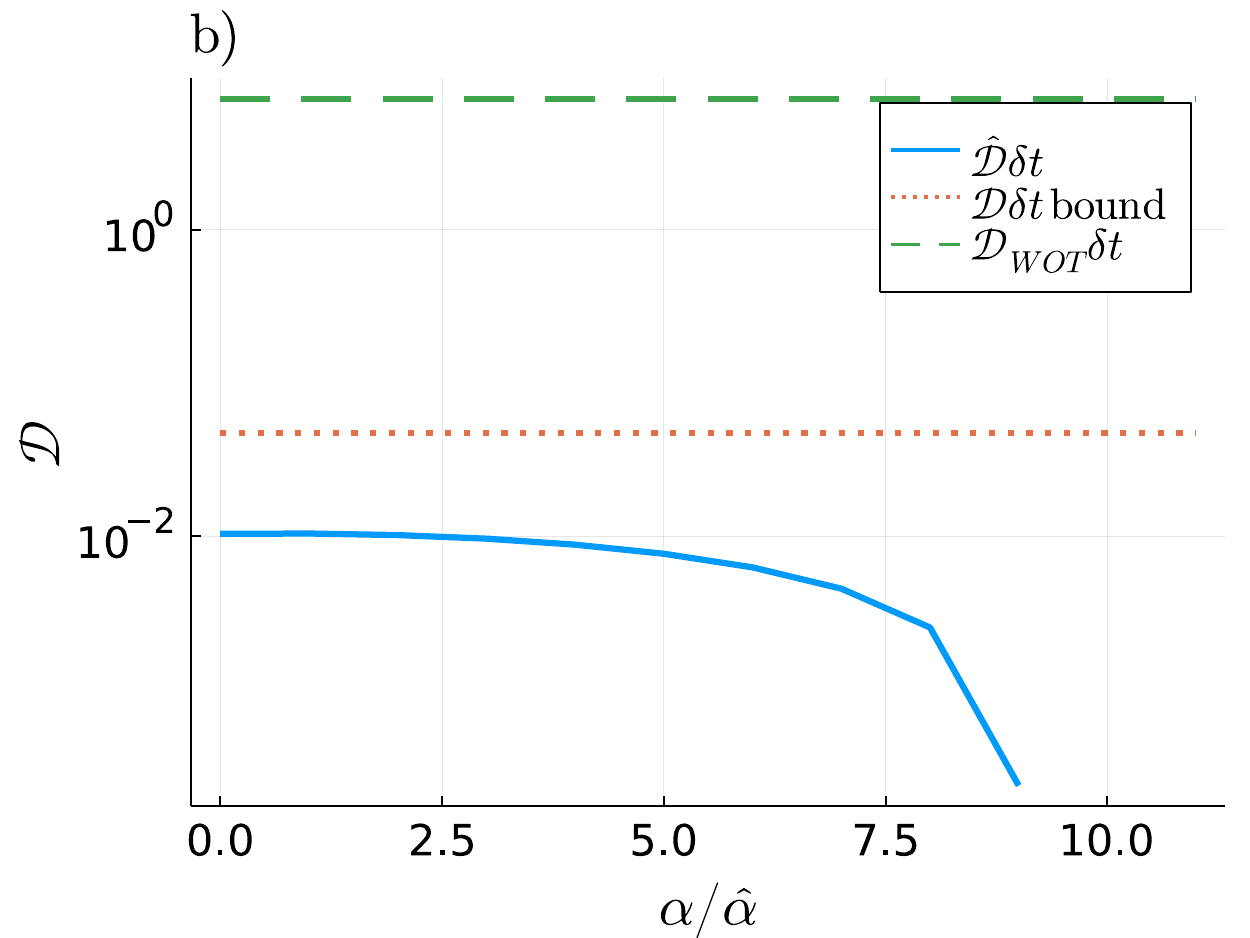}
    \caption{
        Waddington-OT by default incorrectly estimates $\D$ on real data.
        a) A 2d t-SNE~\cite{VanDerMaaten2008} embedding of the hippocampus data from~\cite{LaManno2018}.
        b) The estimate of $\D\popdt$ from the default WOT entropic regularization (dashed green line) is two orders of magnitude larger than $\widehat\D\popdt$ (solid blue line) and the velocity-independent upper bound from Eq.~\eqref{e:covariance_bound_on_D} (dotted red line).
    }
    \label{f:dentate_gyrus}
\end{figure}

Finally, we test $\widehat{D}$ on a publically available dataset on the developing mouse hippocampus. 
La Manno et al.~\cite{LaManno2018} measured gene expression in 8,113 cells at postnatal day 0 (P0) and 10,100 cells at postnatal day 5 (P5).
The dataset is accessible through the \texttt{scvelo} Python package \cite{Bergen2020} as \texttt{dentategyrus\_lamanno}.

Following the standard \texttt{scvelo} preprocessing pipeline, we filter to the top 2,000 highly variable genes, normalize total counts per cell, and log-transform the normalized counts.
Fig.~\ref{f:dentate_gyrus}a shows the data reduced to 2d with t-SNE \cite{VanDerMaaten2008}.
We then compute velocities with \texttt{scvelo}'s dynamical mode, which further reduces the number of genes to 1089.
To calculate $\widehat{D}$, we use the raw velocities produced by \texttt{scvelo}.
$\D_{WOT}\delta t$ is computed after reducing to 50 dimensions with PCA.

Because the velocities are noisy and we do not have an estimate of $\|\ve{\hat v}(\ve{x}) - \ve{v}(\ve{x})\|_\infty$, we are also interested in what can be said about $\D$ from the $\ve{X}(t)$ samples alone.
In Eq.~\eqref{e:covariance_difference_D_estimate}, $\widehat\cov(\ve{X}(\timeone) + \popdt \ve{ \hat v}(\ve{X}(\timeone), \timeone))$ is always positive semidefinite. Hence for any $\ve{\hat v}(\ve{x},t)$,
\begin{align}
    \widehat{\D} \preceq \frac{1}{2\popdt} \cov(\ve{X}(\timetwo)).
    \label{e:covariance_bound_on_D}
\end{align}
From Eq.~\eqref{e:covariance_bound_on_D} we get an estimated upper bound on $\D\popdt$ that does not rely on velocity data.

In the dentate gyrus dataset, $\D_{WOT}\popdt$ is two orders of magnitude larger than both $\widehat{\D}$ and the velocity-independent upper bound $\tr(\cov(\ve{X}(\timetwo)))/2$. 
Fig.~\ref{f:dentate_gyrus}b presents $\widehat{\D}$ for a range of values of the velocity timescale $\alpha$, normalized to the estimate $\hat\alpha$ from Eq.~\eqref{e:velocity_timescale_estimator}.
$\widehat{\D}$ will always be negative for sufficiently large values of $\alpha$, as occurs for $\alpha \ge 10\hat\alpha$ here; in this dataset, $\widehat{\D}$ is relatively insensitive to changes in $\alpha$ near $\hat \alpha$.

A contributing factor in the poor performance of $\D_{WOT}$ in this dataset is that it scales incorrectly with the dimension $d$ of the data.
If each dimension has variability on the same scale, pairwise squared distances are $O(d)$, while the scalar $\tr(\D)/d$ is $O(1)$.
Because of the overestimation of $\D$, running WOT with default settings on this dataset would underestimate the degree to which cell fates are specified at P0 in the dentate gyrus.
In other settings $\D_{WOT}$ may perform better or worse, as its bias depends on $\popdt$ and other features of the data.


\section{Conclusion}
\label{s:conclusion}

Learning single-cell trajectories from destructive measurements is an important challenge for modern biology.
As highlighted by Weinreb et al.~\cite{Weinreb2018}, the data from common experiments is plagued by statistical identifiability problems.
In the years since Weinreb et al.'s paper, progress has been made on many of the obstacles they identified.
The assumption that the population is stationary can be avoided with multiple population measurements~\cite{Lavenant2024,Schiebinger2019,Ventre2023,Wang2022};
 drift fields recovered from RNA velocity~\cite{LaManno2018} or metabolic labeling~\cite{Battich2020} need not be gradients;
 and lineage tracing reveals information about proliferation rates~\cite{Ventre2023}.
The one term in the SDE model (Eq.~\eqref{e:SDE}) which, to the best of our knowledge, no method has attempted to measure directly, is the diffusion matrix $\D(\ve{x}, t)$.
Here, by combining velocity measurements with population time series, we have created a consistent estimator of $\ex{\D(\ve{X}(t), t)}$.

Our estimator $\widehat\D$ has three main disadvantages.
First, it requires accurate estimation of $\ve{v}(\ve{x}, t)$, which is difficult given the strong modeling assumptions and high levels of noise in current velocity inference pipelines~\cite{Gorin2022,Li2020}.
Second, $\widehat{\D}$ is consistent only in the limit $\popdt \to 0$ of closely spaced experimental time points. 
Whether estimators exist that are consistent with finite $\popdt$ remains an open question.
Third, we have only attempted to estimate $\ex{\D(\ve{X}(t), t)}$ rather than the full function $\D(\ve{x}, t)$.
We conjecture that $\D(\ve{x}, t)$ is identifiable in principle, perhaps via stratification and estimating local covariance matrices.
Despite these limitations, $\widehat{\D}$ meaningfully improves on the current state of the art, which makes no inference about $\D$ directly from sequencing data.

\bibliographystyle{plain}
\bibliography{references_diffusion}

\begin{thebibliography}{10}

\bibitem{Battich2020}
Nico Battich, Joep Beumer, Buys {De Barbanson}, Lenno Krenning, Chlo{\'{e}}~S.
  Baron, Marvin~E. Tanenbaum, Hans Clevers, and Alexander {Van Oudenaarden}.
\newblock {Sequencing metabolically labeled transcripts in single cells reveals
  mRNA turnover strategies}.
\newblock {\em Science}, 367(6483):1151--1156, 2020.

\bibitem{Bergen2020}
Volker Bergen, Marius Lange, Stefan Peidli, F.~Alexander Wolf, and Fabian~J.
  Theis.
\newblock Generalizing rna velocity to transient cell states through dynamical
  modeling.
\newblock {\em Nature Biotechnology}, 38(12), 2020.

\bibitem{Brunton2016}
Steven~L. Brunton, Joshua~L. Proctor, and J.~Nathan Kutz.
\newblock {Discovering governing equations from data: Sparse identification of
  nonlinear dynamical systems}.
\newblock {\em PNAS}, 113(15), 2016.

\bibitem{Cavagna2013}
A.~Cavagna, S.~M. {Duarte Queir{\'{o}}s}, I.~Giardina, F.~Stefanini, and
  M.~Viale.
\newblock {Diffusion of individual birds in starling flocks}.
\newblock {\em Proceedings of the Royal Society B: Biological Sciences},
  280(1756), 2013.

\bibitem{Chen2022}
Yongxin Chen, Tryphon~T. Georgiou, and Michele Pavon.
\newblock The most likely evolution of diffusing and vanishing particles:
  Schrödinger bridges with unbalanced marginals.
\newblock {\em SIAM Journal on Control and Optimization}, 60(4):2016--2039,
  2022.

\bibitem{Forrow2021}
Aden Forrow and Geoffrey Schiebinger.
\newblock {LineageOT is a Unified Framework for Lineage Tracing and Trajectory
  Inference}.
\newblock {\em Nature Communications}, 12:4940, 2021.

\bibitem{Frishman2020}
Anna Frishman and Pierre Ronceray.
\newblock {Learning Force Fields from Stochastic Trajectories}.
\newblock {\em Physical Review X}, 10(2):21009, 2020.

\bibitem{Gardiner2004}
C.~W Gardiner.
\newblock {\em Handbook of stochastic methods : for physics, chemistry and the
  natural sciences}.
\newblock Springer series in synergetics (Unnumbered). Springer-Verlag, Berlin,
  3rd ed. edition, 2004.

\bibitem{Gorin2022}
Gennady Gorin, Meichen Fang, Tara Chari, and Lior Pachter.
\newblock {RNA velocity unraveled}.
\newblock {\em PLoS Computational Biology}, 18(9):1--55, 2022.

\bibitem{LaManno2018}
Gioele {La Manno}, Ruslan Soldatov, Amit Zeisel, Emelie Braun, Hannah
  Hochgerner, Viktor Petukhov, Katja Lidschreiber, Maria~E Kastriti, Peter
  L{\"{o}}nnerberg, Alessandro Furlan, Jean Fan, Lars~E Borm, Zehua Liu, David
  van Bruggen, Jimin Guo, Xiaoling He, Roger Barker, Erik Sundstr{\"{o}}m,
  Gon{\c{c}}alo Castelo-Branco, Patrick Cramer, Igor Adameyko, Sten Linnarsson,
  and Peter~V Kharchenko.
\newblock {RNA velocity of single cells.}
\newblock {\em Nature}, 560(7719):494--498, 2018.

\bibitem{Lange2022}
Marius Lange, Volker Bergen, Michal Klein, Manu Setty, Bernhard Reuter, Mostafa
  Bakhti, Heiko Lickert, Meshal Ansari, Janine Schniering, Herbert~B. Schiller,
  Dana Pe'er, and Fabian~J. Theis.
\newblock {CellRank for directed single-cell fate mapping}.
\newblock {\em Nature Methods}, 19(2):159--170, 2022.

\bibitem{Lavenant2024}
Hugo Lavenant, Stephen Zhang, Young~Heon Kim, and Geoffrey Schiebinger.
\newblock {Toward a Mathematical Theory of Trajectory Inference}.
\newblock {\em Annals of Applied Probability}, 34(1):428--500, 2024.

\bibitem{Li2020}
Tiejun Li, Jifan Shi, Yichong Wu, and Peijie Zhou.
\newblock {On the Mathematics of RNA Velocity I: Theoretical Analysis}.
\newblock {\em bioRxiv preprint}, 2020.

\bibitem{Qiu2017}
Xiaojie Qiu, Qi~Mao, Ying Tang, Li~Wang, Raghav Chawla, Hannah~A. Pliner, and
  Cole Trapnell.
\newblock {Reversed graph embedding resolves complex single-cell trajectories}.
\newblock {\em Nature Methods}, 14(10):979--982, 2017.

\bibitem{Qiu2022}
Xiaojie Qiu, Yan Zhang, Shayan Hosseinzadeh, Dian Yang, Angela Pogson, Li~Wang,
  Matt Shurtleff, Ruoshi Yuan, Song Xu, Yian Ma, Joseph Replogle, Spyros
  Darmanis, Ivet Bahar, Jianhua Xing, and Jonathan Weissman.
\newblock {Mapping Transcriptomic Vector Fields of Single Cells}.
\newblock {\em Cell}, 2022.

\bibitem{Schiebinger2019}
Geoffrey Schiebinger, Jian Shu, Marcin Tabaka, Brian Cleary, Vidya Subramanian,
  Aryeh Solomon, Joshua Gould, Siyan Liu, Stacie Lin, Peter Berube, Lia Lee,
  Jenny Chen, Justin Brumbaugh, Philippe Rigollet, Konrad Hochedlinger, Rudolf
  Jaenisch, Aviv Regev, and Eric~S Lander.
\newblock {Optimal-Transport Analysis of Single-Cell Gene Expression Identifies
  Developmental Trajectories in Reprogramming}.
\newblock {\em Cell}, 176(4):928----943.e22, 2019.

\bibitem{Street2018}
Kelly Street, Davide Risso, Russell~B Fletcher, Diya Das, John Ngai, Nir Yosef,
  Elizabeth Purdom, and Sandrine Dudoit.
\newblock {Slingshot: cell lineage and pseudotime inference for single-cell
  transcriptomics.}
\newblock {\em BMC genomics}, 19(1):477, 2018.

\bibitem{Stuart1994}
Alan Stuart and Keith Ord.
\newblock {\em {Kendall's Advanced Theory of Statistics}}, volume 1
  (Distribution Theory).
\newblock Halsted Press, 1994.

\bibitem{VanDerMaaten2008}
L~J~P {Van Der Maaten} and G~E Hinton.
\newblock {Visualizing data using t-SNE}.
\newblock {\em Journal of Machine Learning Research}, 9:2579--2605, 2008.

\bibitem{Ventre2023}
Elias Ventre, Aden Forrow, Nitya Gadhiwala, Parijat Chakraborty, Omer Angel,
  and Geoffrey Schiebinger.
\newblock {Trajectory inference for a branching SDE model of cell
  differentiation}.
\newblock 2023.

\bibitem{Wang2022}
Shou-Wen Wang, Michael~J. Herriges, Kilian Hurley, Darrell~N. Kotton, and
  Allon~M. Klein.
\newblock {CoSpar identifies early cell fate biases from single-cell
  transcriptomic and lineage information}.
\newblock {\em Nature Biotechnology}, 40:1066--1074, 2-22.

\bibitem{Weinreb2018}
Caleb Weinreb, Samuel Wolock, Betsabeh~K. Tusi, Merav Socolovsky, and Allon~M.
  Klein.
\newblock {Fundamental limits on dynamic inference from single-cell snapshots}.
\newblock {\em Proceedings of the National Academy of Sciences of the United
  States of America}, 115(10):E2467--E2476, 2018.

\bibitem{Zhang2021}
Stephen Zhang, Anton Afanassiev, Laura Greenstreet, Tetsuya Matsumoto, and
  Geoffrey Schiebinger.
\newblock {Optimal transport analysis reveals trajectories in steady-state
  systems}.
\newblock {\em PLoS Computational Biology}, 17(12):1--29, 2021.

\end{thebibliography}

\section{Supplement}
\subsection{Code availability}
Code to reproduce figures in this paper is available at \url{https://github.com/aforr/diffusion_estimation}.

\subsection{Proof of Theorem~\ref{thm:consistency}}
\label{s:convergence_proof}

Here we prove a variant of Theorem~\ref{thm:consistency} with explicit bounds.
We begin by laying out in Section~\ref{appendix:proof_assumptions} the moment assumptions we will need.
Altogether, these amount to assuming $\ve{X}(t)$, $\ve{v}(\ve{X}(t), t)$, and $\D(\ve{X}(t), t)$ have bounded fourth moments, the derivatives of $\ve{v}(\ve{X}(t), t)$ and $\D(\ve{X}(t), t)$ to first order in $t$ and second order in $\ve{x}$ have bounded second moments, and the error in the velocity estimate $\ve{\hat v}(\ve{x}, t)$ is bounded.
Throughout, $\ve{v}(\ve{x}, t)$ and $\D(\ve{x}, t)$ are fixed. The only randomness comes from $\ve{X}(t)$.

In Section~\ref{appendix:theorem_statement} we state our main result, Theorem~\ref{thm:consistency_long}, and prove it with reference to later lemmas.
We also prove a lemma bounding all of the bias terms in $\widehat{\D}$. 
In Sections~\ref{appendix:sampling_error_lemma}, \ref{appendix:time_discretization_lemma}, and \ref{appendix:velocity_error_lemma}, respectively, we prove lemmas bounding the error in $\widehat{\D}$ from sampling error, finite $\popdt$, and imperfect velocity estimation.

We have not optimized the constant factors and expect the analysis could be tightened to give a sharper bound.

\subsubsection{Assumptions}
\label{appendix:proof_assumptions}
\begin{assumption}
    The central and noncentral fourth moments of $\ve{X}(t)$ and $\ve{Y}(t) = \ve{X}(\timeone) + \popdt \ve{\hat v}(\ve{X}(t), t)$ are bounded for all $t$ by $C_{\ve{x}}^4$ with $C_{\ve{x}}\ge 0$, i.e. for $i = 1, \ldots d$,
    \begin{align}
        C_{\ve{x}}^4 &\ge \max\bigg( \ex{X_i(t)^4}, \ex{(X_i(t) - \ex{X_i(t)})^4}, \nonumber
        \\
        & \qquad\qquad\,
        \ex{Y_i(t)^4}, \ex{(Y_i(t) - \ex{Y_i(t)})^4} \bigg).
    \end{align}
    \label{assumption:x_bound}
\end{assumption}
By Jensen's inequality, Assumption~\ref{assumption:x_bound} also implies 
\begin{align}
    C_{\ve{x}}^2 &\ge \ex{X_i(t)^2} \ge \var(X_i(t)),
    \\
    C_{\ve{x}} & \ge | \ex{X_i(t)}|,
\end{align}
and similar inequalities for $Y_i$.

\begin{assumption}
    For all $t$ and $i = 1, \ldots d$,
    \begin{align}
        C_{\ve{v}}^4 &\ge \ex{v_i(\ve{X}(t),t)^4},
    \end{align}
    with $C_{\ve{v}}\ge 0$.
    \label{assumption:velocity_bound}
\end{assumption}

Again Jensen's inequality gives us similar bounds on lower moments from Assumption~\ref{assumption:velocity_bound}:
\begin{align}
    C_{\ve{v}}^2 & \ge \ex{v_i(\ve{X}(t),t)^2} \ge \var(v_i(\ve{X}(t),t)),
    \\
    C_{\ve{v}} & \ge |\ex{v_i(\ve{X}(t),t)}|.
\end{align}
The same calculations apply for Assumptions~\ref{assumption:velocity_derivative_bounds}-\ref{assumption:D_derivative_bounds}.

\begin{assumption}
    The derivatives of $\ve{v}(\ve{x}, t)$ to first order in $t$ and second order in $\ve{x}$ have bounded second moments, i.e. for all $t$ and $i = 1, \ldots d$
    \begin{align}
        C_{\ve{v}_t}^2 &\ge \ex{\left(\frac{\partial}{\partial t} v_i(\ve{X}(t),t)\right)^2} ,
        \\
        C_{\ve{v}_x}^2 & \ge \ex{\left(\frac{\partial}{\partial x_k}v_i(\ve{X}(t),t)\right)^2},
        \\
        C_{\ve{v}_{xx}}^2 & \ge \ex{\left(\frac{\partial^2}{\partial x_k \partial_\ell} v_i(\ve{X}(t),t)\right)^2} .
    \end{align}
    \label{assumption:velocity_derivative_bounds}
\end{assumption}

\begin{assumption}
    For all $t$ and $i, j = 1, \ldots d$,
    \begin{align}
        C_{\D}^4 \ge \ex{\D_{ij}(\ve{X}(t),t)^4}
    \end{align}
    with $C_{\D} \ge 0$.
    \label{assumption:D_bound}
\end{assumption}

\begin{assumption}
    The derivatives of $\D(\ve{x}, t)$ to first order in $t$ and second order in $\ve{x}$ have bounded second moments, i.e. for all $t$ and $i, j = 1, \ldots d$
    \begin{align}
        C_{\D_t}^2 &\ge \ex{\left(\frac{\partial}{\partial t} \D_{ij}(\ve{X}(t),t)\right)^2},
        \\
        C_{\D_x}^2 &\ge \ex{\left(\frac{\partial}{\partial x_k} \D_{ij}(\ve{X}(t),t)\right)^2},
        \\
        C_{\D_{xx}}^2 & \ge \ex{\left(\frac{\partial^2}{\partial x_k \partial x_\ell}  \D_{ij}(\ve{X}(t),t)\right)^2}.
    \end{align}
    \label{assumption:D_derivative_bounds}
\end{assumption}

The final assumption is a bound on the error of our regression estimate of the drift function.
\begin{assumption}
    For all $\ve{x}$ and $t$,
    \begin{align}
        C_{{\ve{\hat v} - \ve{v}}} \ge \|\ve{\hat v}(\ve{x}, t) - \ve{v}(\ve{x}, t)\|_\infty.
    \end{align}
    \label{assumption:v_error_bound}
\end{assumption}

\subsubsection{Theorem statement}
\label{appendix:theorem_statement}
\begin{theorem}
    Suppose $p(\ve{x}, t)$ is a solution to Eq.~\eqref{e:Fokker-Planck} such that Assumptions~\ref{assumption:x_bound}-\ref{assumption:D_derivative_bounds} hold with $\ve{X}(t)\sim p(\ve{x}, t)$. Let $\errprob > 0$.
    
    Given $n\ge 2$ iid samples from $p(\ve{x}, t)$, $n$ iid samples from $p(\ve{x}, t+ \popdt)$, and a regression estimate $\ve{\hat v}(\ve{x}, t)$ satisfying Assumption~\ref{assumption:v_error_bound}, with probability at least $1 - \errprob$
    \begin{align}
        \|\widehat{\D} - \ex{\D(\ve{X}(t), t)} \|_\infty \le C_{\ve{\hat v} - \ve{v}}C_{\ve{x}} + \frac{1}{2}C_{\ve{\hat v} - \ve{v}}^2 \delta t + \left(\frac{C}{4} + \frac{C_{\ve{v}}^2}{2}\right) \delta t + \frac{d C_{\ve{x}}^2}{ \delta t \sqrt{n\errprob}},
        \label{e:D_error_bound_explicit}
    \end{align}
    where $C$ is given by Eq.~\eqref{e:constant_in_time_discretization_error}.
    \label{thm:consistency_long}
\end{theorem}

\begin{proof}
    We start by separating out the bias in $\widehat{\D}$:
    \begin{align}
        \|\widehat{\D} - \ex{\D(\ve{X}(t), t)} \|_\infty & = \left\|\widehat{\D} - \ex{\widehat\D} + \ex{\widehat{\D}}- \ex{\D(\ve{X}(t), t)} \right\|_\infty
        \\
        &\le  \left\|\widehat{\D} - \ex{\widehat\D} \right\|_\infty + \left\|\ex{\widehat{\D}}- \ex{\D(\ve{X}(t), t)} \right\|_\infty.
        \label{e:bias_sampling_error_split}
    \end{align}
    From Lemma~\ref{lemma:sampling_error}, with probability at least $1-\errprob$
    \begin{align}
        \left\|\widehat{\D} - \ex{\widehat\D} \right\|_\infty \le \frac{d C_{\ve{x}}^2}{ \delta t \sqrt{n\errprob}}.
        \label{e:proof_sampling_bound}
    \end{align}
    
    From Lemma~\ref{lemma:joint_bias_bound},
    \begin{align}
        \left\|\ex{\widehat{\D}}- \ex{\D(\ve{X}(t), t)} \right\|_\infty \le C_{\ve{\hat v} - \ve{v}}C_{\ve{x}} + \frac{1}{2}C_{\ve{\hat v} - \ve{v}}^2 \delta t + \left(\frac{C}{4} + \frac{C_{\ve{v}}^2}{2}\right) \delta t.
        \label{e:proof_bias_bound}
    \end{align}
    Combining Eqns.~\eqref{e:proof_sampling_bound} and \eqref{e:proof_bias_bound} with Eq.~\eqref{e:bias_sampling_error_split} yields Eqn.~\eqref{e:D_error_bound_explicit}.
\end{proof}

Conditional on the velocity estimate $\ve{\hat v}$, there are two sources of bias: error in the velocity estimate and error from finite $\delta t$.
\begin{lemma}[Bias bound]
    If Assumptions~\ref{assumption:x_bound}-\ref{assumption:v_error_bound} hold
    and 
    $C$ is given by Eq.~\eqref{e:constant_in_time_discretization_error},
    then
    \begin{align}
        \left\|\ex{\widehat{\D}}- \ex{\D(\ve{X}(t), t)} \right\|_\infty
        \le C_{\ve{\hat v} - \ve{v}}C_{\ve{x}} + \frac{1}{2}C_{\ve{\hat v} - \ve{v}}^2 \delta t + \left(\frac{C}{4} + \frac{C_{\ve{v}}^2}{2}\right) \delta t.
    \end{align}
    \label{lemma:joint_bias_bound}
\end{lemma}
\begin{proof}
    We will first separate out the bias due to imperfect estimation of $\ve{v}(\ve{x}, t)$. Define $\widehat{\D}_{\ve{v}}$ by replacing $\ve{\hat v}(\ve{x}, t)$ in $\widehat{\D}$ with the unknown true velocity $\ve{v}(\ve{x}, t)$, i.e.
    \begin{align}
        \widehat{\D}_{\ve{v}} & = \frac{1}{2\popdt}\left(\widehat\cov(\ve{X}(\timetwo)) - \widehat\cov(\ve{X}(\timeone) + \popdt \ve{v}(\ve{X}(\timeone), \timeone))\right).
        \label{e:D_estimate_true_velocity}
    \end{align}
    Then
    \begin{align}
        \left\|\ex{\widehat{\D}}- \ex{\D(\ve{X}(t), t)} \right\|_\infty \le \left\|\ex{\widehat{\D}}- \ex{\widehat{\D}_{\ve{v}}} \right\|_\infty + \left\|\ex{\widehat{\D}_{\ve{v}}}- \ex{\D(\ve{X}(t), t)} \right\|_\infty.
        \label{e:bias_split_in_proof}
    \end{align}
    For the first term,
    \begin{align}
        \ex{\widehat{\D} - \widehat{\D}_\ve{v}} & = \frac{1}{2\delta t}\left(\cov(\ve{X}(t) + \delta t \ve{\hat v}(\ve{X}(t), t)) - \cov(\ve{X}(t) + \delta t \ve{v}(\ve{X(t)}, t))\right).
        \label{e:velocity_error_in_bias_proof}
    \end{align}
    Eq.~\eqref{e:velocity_error_lemma} from Lemma~\ref{lemma:velocity_error} bounds the right side of Eq.~\ref{e:velocity_error_in_bias_proof}, yielding
    \begin{align}
        \ex{\widehat{\D} - \widehat{\D}_\ve{v}} & \le C_{\ve{\hat v} - \ve{v}}C_{\ve{x}} + \frac{1}{2}C_{\ve{\hat v} - \ve{v}}^2 \delta t.
        \label{e:velocity_error_bound_in_bias_proof}
    \end{align}

    To bound $\left\|\ex{\widehat{\D}_{\ve{v}}}- \ex{\D(\ve{X}(t), t)} \right\|_\infty$ we first replace the covariance matrix at $\timetwo = \timeone + \delta t$ with a Taylor series approximation:
    \begin{align}
        \left\|\ex{\widehat{\D}_{\ve{v}}}- \ex{\D(\ve{X}(t), t)} \right\|_\infty & = \left\|\frac{1}{2\delta t}\left(\cov(\ve{X}(\timetwo)) - \cov(\ve{X}(\timeone) + \popdt \ve{v}(\ve{X}(\timeone), \timeone))\right)- \ex{\D(\ve{X}(\timeone), \timeone)} \right\|_\infty
        \\
        & = \frac{1}{2\delta t}\bigg\|\cov(\ve{X}(\timetwo)) - \cov(\ve{X}(t)) - \delta t \ddt (\cov(\ve{X}(t)))\nonumber
        \\
        & \qquad\quad + \cov(\ve{X}(t)) + \delta t \ddt (\cov(\ve{X}(t)))\nonumber
        \\
        & \qquad\quad - \cov(\ve{X}(\timeone) + \popdt \ve{v}(\ve{X}(\timeone), \timeone))- 2\delta t\ex{\D(\ve{X}(t), t)} \bigg\|_\infty
        \\
        & \le \frac{1}{2\delta t}\left\|\cov(\ve{X}(\timetwo)) - \cov(\ve{X}(t)) - \delta t \ddt (\cov(\ve{X}(t)))\right\|_\infty\nonumber
        \\
        & \quad + \frac{1}{2\delta t}\bigg\|\cov(\ve{X}(t)) + \delta t \ddt (\cov(\ve{X}(t)))\nonumber
        \\
        & \qquad\quad - \cov(\ve{X}(\timeone) + \popdt \ve{v}(\ve{X}(\timeone), \timeone))- 2\delta t\ex{\D(\ve{X}(t), t)} \bigg\|_\infty.
        \label{e:bias_bound_with_derivative}
    \end{align}

    For the second term in Eq.~\eqref{e:bias_bound_with_derivative} we need to compute the time derivative of $\cov(\ve{X}(t))$.
    Let $\mathcal{L}_t$ be the generator of the Markov process of Eq.~\eqref{e:SDE}, so that
    \begin{align}
        \mathcal{L}_t f(\ve{x}) = (\nabla f(\ve{x}))^\top \ve{v}(\ve{x}, t) + \trace\left(\D(\ve{x}, t) \nabla^2 f(\ve{x})\right)
        \label{e:generator_definition}
    \end{align}
    and
    \begin{align}
        \ddt p(\ve{x}, t) = \mathcal{L}^*_t p(\ve{x}, t).
    \end{align}
    We can then write
    \begin{align}
        \ddt \cov(\ve{X}(t)) & = \ddt \left(\mathbb{E}[\ve{X}(t)\ve{X}(t)^\top] - \mathbb{E}[\ve{X}(t)]\mathbb{E}[\ve{X}(t)]^\top\right) 
        \\
        & = \mathbb{E}[\mathcal{L}_t(\ve{X}(t)\ve{X}(t)^\top)] - \mathbb{E}[\mathcal{L}_t\ve{X}(t)]\mathbb{E}[\ve{X}(t)]^\top - \mathbb{E}[\ve{X}(t)]\mathbb{E}[\mathcal{L}_t\ve{X}(t)]^\top.
        \label{e:covariance_derivative_with_L}
    \end{align}
    Applying Eq.~\eqref{e:generator_definition},
    \begin{align}
        \mathcal{L}_t(x_i x_j) & = \left(\sum_k v_k(\ve{x}, t) (\delta_{ki} x_j + x_i \delta_{kj})\right) + \sum_{k\ell} \D_{k\ell}(\delta_{ki}\delta_{\ell j}+\delta_{\ell i}\delta_{kj})
        \\
        \mathcal{L}_t(\ve{x}\ve{x}^\top) & = \ve{v}(\ve{x}, t)\ve{x}^\top + \ve{x} \ve{v}(\ve{x}, t)^\top + 2 \D(\ve{x}, t)
        \label{e:L_xxT}
        \\
        \mathcal{L}_t(x_i) & = \sum_{k} v_k(\ve{x}, t) \delta_{ki}x_i
        \\
        \mathcal{L}_t(\ve{x}) & = \ve{v}(\ve{x}, t).
        \label{e:L_x}
    \end{align}
    Substituting Eqns.~\eqref{e:L_xxT} and~\eqref{e:L_x} into Eq.~\eqref{e:covariance_derivative_with_L}, we find
    \begin{align}
        \ddt(\cov(\ve{x})) & = \mathbb{E}[2\D(\ve{X}(t), t) + \ve{v}(\ve{X}(t), t)\ve{X}(t)^\top + \ve{X}(t) \ve{v}(\ve{X}(t), t)^\top]
        \nonumber 
        \\
        & \quad - \ex{\ve{v}(\ve{X}(t), t)}\ex{\ve{X}(t)^\top} - \ex{\ve{X}(t)}\ex{\ve{v}(\ve{X}(t), t)^\top}.
    \end{align}

    Returning to Eq.~\eqref{e:bias_bound_with_derivative}, we next expand the pushforward covariance matrix:
    \begin{align}
        \cov(\ve{X}(\timeone) + \delta t \ve{v}(\ve{X}(\timeone), \timeone)) & = \cov(\ve{X}(\timeone)) \nonumber
        \\
        & \quad + \delta t \ex{\ve{v}(\ve{X}(\timeone), \timeone)\ve{X}(\timeone)^\top + \ve{X}(\timeone) \ve{v}(\ve{X}\timeone, \timeone)^\top}\nonumber
        \\
        & \quad - \delta t \left(\ex{\ve{v}(\ve{X}(\timeone), \timeone)}\ex{\ve{X}(\timeone)^\top} + \ex{\ve{X}(\timeone)}\ex{\ve{v}(\ve{X}(\timeone), \timeone)^\top}\right)\nonumber
        \\
        & \quad + \delta t^2 \cov(\ve{v}(\ve{X}(\timeone), \timeone))
        \\
        & = \cov(\ve{X}(\timeone)) + \delta t \ddt (\cov(\ve{X}(\timeone))) - 2\delta t\ex{\D(\ve{X}(\timeone), \timeone)} + \delta t^2 \cov(\ve{v}(\ve{X}(\timeone), \timeone)).
        \label{e:pushforward_covariance_expansion}
    \end{align}
    Most of the terms in Eq.~\eqref{e:pushforward_covariance_expansion} cancel when inserted in Eq.~\eqref{e:bias_bound_with_derivative}, leaving
    \begin{align}
        \left\|\ex{\widehat{\D}_{\ve{v}}}- \ex{\D(\ve{X}(t), t)} \right\|_\infty & \le \frac{1}{2\delta t}\left\|\cov(\ve{X}(\timetwo)) - \cov(\ve{X}(t)) - \delta t \ddt (\cov(\ve{X}(t)))\right\|_\infty
        + \frac{1}{2\delta t}\left\|\delta t^2 \cov(\ve{v}(\ve{X}(\timeone), \timeone))\right\|_\infty.
    \end{align}

    The first term is bounded by $C\delta t/4$ by Lemma~\ref{lemma:covariance_Taylor_bound}. For the second term, using Assumption~\ref{assumption:velocity_bound} and the Cauchy-Schwarz inequality,
    \begin{align}
        |\cov(v_i(\ve{X}(\timeone), \timeone), v_j(\ve{X}(\timeone), \timeone))| & \le \var(v_i(\ve{X}(\timeone), \timeone))^{1/2}\var(v_j(\ve{X}(\timeone), \timeone))^{1/2}
        \\
        & \le C_{\ve{v}}^2.
    \end{align}
    Hence
    \begin{align}
        \left\|\ex{\widehat{\D}_{\ve{v}}}- \ex{\D(\ve{X}(t), t)} \right\|_\infty \le \left(\frac{C}{4} + \frac{C_{\ve{v}}^2}{2}\right) \delta t.
    \end{align}
    Combining with Eqns.~\eqref{e:bias_split_in_proof} and \eqref{e:velocity_error_bound_in_bias_proof} leads to
    \begin{align}
        \left\|\ex{\widehat{\D}}- \ex{\D(\ve{X}(t), t)} \right\|_\infty
        \le  C_{\ve{\hat v} - \ve{v}}C_{\ve{x}} + \frac{1}{2}C_{\ve{\hat v} - \ve{v}}^2 \delta t + \left(\frac{C}{4} + \frac{C_{\ve{v}}^2}{2}\right) \delta t.
    \end{align}
    as desired.
\end{proof}

\subsubsection{Sampling error}
\label{appendix:sampling_error_lemma}
\begin{lemma}[$\widehat{\D}$ sampling error]
    Suppose that Assumption~\ref{assumption:x_bound} holds and $n\ge 2$.
    Then with probability at least $1-\errprob$
    \begin{align}
        \left\|\widehat{\D} - \ex{\widehat\D} \right\|_\infty < \frac{d C_{\ve{x}}^2}{ \delta t \sqrt{n\errprob}}.
    \end{align}
    \label{lemma:sampling_error}
\end{lemma}
\begin{proof}
    We start by relating the matrix norm to the size of the entries with a union bound. For any $\zeta\in \mathbb{R}$,
    \begin{align}
        P\left(\left\|\widehat{\D} - \ex{\widehat\D} \right\|_\infty \ge \zeta\right) & \le \sum_{ij}  P\left(\left|\widehat\D_{ij} - \ex{\widehat\D_{ij}}\right| \ge \zeta\right).
        \label{e:sampling_error_union_bound}
    \end{align}
    By Chebyshev's inequality,
    \begin{equation}
        P\left(\left|\widehat\D_{ij} - \ex{\widehat\D_{ij}}\right| \ge \zeta\right) \le \frac{\var(\widehat\D_{ij})}{\zeta^2}.
        \label{e:sampling_error_chebyshev}
    \end{equation}
    Since our samples at $t$ and $t+\delta t$ are independent,
    \begin{align}
        \var\left(\widehat{\D_{ij}}\right) & = \frac{1}{4\delta t^2}\left( \var\left(\widehat\cov(\ve{X}(\timetwo))_{ij}\right) + \var\left(\widehat\cov(\ve{X}(\timeone) + \popdt \ve{ \hat v}(\ve{X}(\timeone), \timeone))_{ij}\right)\right).
    \end{align}
    Both of the variance terms on the right are bounded by Lemma~\ref{lemma:variance_of_covariance_bound}. Hence
    \begin{align}
        \var\left(\widehat{\D_{ij}}\right) & \le \frac{1}{4\delta t^2} \left( \frac{2C_{\ve{x}}^4}{n} + \frac{2C_{\ve{x}}^4}{n}\right)
        \\
        & = \frac{C_{\ve{x}}^4}{n \delta t^2}.
    \end{align}
    Substituting into Eqns.~\eqref{e:sampling_error_chebyshev} and \eqref{e:sampling_error_union_bound}, we have
    \begin{align}
        P\left(\left|\widehat\D_{ij} - \ex{\widehat\D_{ij}}\right| \ge \zeta\right) \le \frac{C_{\ve{x}}^4}{n \delta t^2\zeta^2}
    \end{align}
    and
    \begin{align}
        P\left(\left\|\widehat{\D} - \ex{\widehat\D} \right\|_\infty \ge \zeta\right) & \le \frac{d^2C_{\ve{x}}^4}{n \delta t^2\zeta^2}.
    \end{align}
    We conclude by setting
    \begin{align}
        \errprob = \frac{d^2C_{\ve{x}}^4}{n \delta t^2\zeta^2}
    \end{align}
    so that
    \begin{align}
        \zeta = \left(
            \frac{d^2C_{\ve{x}}^4}{n \delta t^2 \errprob}
        \right)^{1/2}.
    \end{align}
\end{proof}

\begin{lemma}[Bound on variance of covariance]
    If $n\ge 2$ and $\ex{(X_i - \ex{X_i})^4} \le C_{\ve{x}}^{4}$ for $i=1, \ldots d$, then
    \begin{align}
        \var(\widehat{\cov}(\ve{X})_{ij}) \le \frac{2C_{\ve{x}}^4}{n}.
    \end{align}
    \label{lemma:variance_of_covariance_bound}
\end{lemma}
\begin{proof}
    From Lemma~\ref{lemma:variance_of_covariance},
    \begin{align}
        \var(\widehat{\cov}(\ve{X})_{ij}) & = \frac{\ex{(X_i - \ex{X_i})^2(X_j - \ex{X_j})^2}}{n} -\frac{(n-2)\cov(X_i,X_j)^2}{n(n-1)} + \frac{\var(X_i)\var(X_j)}{n(n-1)}
        \\
        & \le \frac{\ex{(X_i - \ex{X_i})^2(X_j - \ex{X_j})^2}}{n} + \frac{\var(X_i)\var(X_j)}{n(n-1)}.
    \end{align}
    The fourth moment assumption bounds the first term by the Cauchy-Schwarz inequality and the second term by Jensen's inequality:
    \begin{align}
        \ex{(X_i - \ex{X_i})^2(X_j - \ex{X_j})^2} & \le \ex{(X_i - \ex{X_i})^4}^{1/2}\ex{(X_j - \ex{X_j})^4}^{1/2} \le C_{\ve{x}}^4
        \\
        \var(X_i) &\le \ex{(X_i - \ex{X_i})^4}^{1/2} \le C_{\ve{x}}^2.
    \end{align}
    Since the numerators of both fractions are less than or equal to $C_{\ve{x}}^4$,
    \begin{align}
        \var(\widehat{\cov}(\ve{X})_{ij}) & \le \frac{C_{\ve{x}}^4}{n} \left(1 + \frac{1}{n-1}\right)
        \\
        & \le \frac{2C_{\ve{x}}^4}{n}.
    \end{align}
\end{proof}
\begin{lemma}[Variance of empirical covariance]
    Given $n\ge 2$ iid samples $(X_i, Y_i)$, $i = 1, \ldots, n$, 
    \begin{align}
        \var\left(
            \frac{1}{n-1}\sum_{i=1}^n (X_i - \bar X)(Y_i - \bar Y)
        \right) = \frac{\ex{(X - \ex{X})^2(Y - \ex{Y})^2}}{n} -\frac{(n-2)\cov(X,Y)^2}{n(n-1)} + \frac{\var(X)\var(Y)}{n(n-1)}.
        \label{e:variance_of_covariance}
    \end{align}
    \label{lemma:variance_of_covariance}
\end{lemma}
The proof of Lemma~\ref{lemma:variance_of_covariance} involves extensive algebra. The necessary algebraic techniques are presented in Chapter 13 of \cite{Stuart1994}. 
In their notation, the quantity we seek to compute is $\var(k_{11})$, where $k_{11}$ is given by Eq. (13.2). 
Although Stuart and Ord do not explicitly write out our Eq.~\eqref{e:variance_of_covariance}, it can be reconstructed by noting $\ex{k_{11}} = \kappa_{11}$, using Eq. (13.15) for $\ex{k_{11}^2}$, and applying Eqs. (3.81) to translate the cumulants $\kappa$ into the moments we use in Lemma~\ref{lemma:variance_of_covariance}.

\subsubsection{Time discretization error}
\label{appendix:time_discretization_lemma}
\begin{lemma}[Time-discretization error]
    If Assumptions~\ref{assumption:x_bound}-\ref{assumption:D_derivative_bounds} hold, then 
    \begin{align}
        \left\|\cov(\ve{X}(t + \delta t)) - \cov(\ve{X}(t)) - \delta t \ddt (\cov(\ve{X}(t)))\right\|_\infty \le \frac{C}{2} \delta t^2,
    \end{align}
    where
    \begin{align}
        C = 2\bigg(2C_{\ve{v}}^2 + 2d C_{\ve{x}} C_{\ve{v}} C_{\ve{v}_x} + dC_{\ve{v}} C_{\D_x} + 2d^2 C_{\ve{x}}C_{\D} C_{\ve{v}_{xx}}
        + 2 d C_{\D} C_{\ve{v}_x} +  d^2 C_{\D} C_{\D_{xx}} +  2C_{\ve{x}} C_{\ve{v}_t} + C_{\D_t}\bigg).
        \label{e:constant_in_time_discretization_error}
    \end{align}
    \label{lemma:covariance_Taylor_bound}
\end{lemma}

\begin{proof}
    To simplify and clarify expressions in this proof, we will suppress the arguments of $p(\ve{x}, t)$, $\ve{v}(\ve{x}, t)$, and $\D(\ve{x}, t)$, adopt the Einstein convention where repeated indices are summed, and write derivatives with a single subscript. That is,
    \begin{align}
        \partial_t & = \frac{\partial}{\partial t}
        \\
        \partial_i & = \frac{\partial}{\partial x_i}.
    \end{align}
    In this notation, the generator corresponding to Eq.~\eqref{e:Fokker-Planck} is
    \begin{align}
        \mathcal{L}_t f = v_i\partial_i  f + \D_{ij}\partial_i\partial_j f.
    \end{align}

    Our goal is to prove a bound on $\left\|\partial_t^2 \cov(\ve{X}(t))\right\|_\infty$ and apply Lemma~\ref{lemma:matrix_Taylors_theorem}.
    Since the covariance involves a combination of expectations, we first expand the derivative:
    \begin{align}
        \left|\partial_t^2 \left(\mathbb{E}[X_iX_j] - \ex{X_i}\ex{X_j}^\top\right)\right| & 
        = \bigg|\partial_t^2\ex{X_iX_j}  \nonumber
        \\
        &\quad - \bigg((\partial_t^2\ex{X_i})\ex{X_j}+ \ex{X_i}\partial_t^2\ex{X_j} +        \nonumber
        \\
        & \qquad\quad 2 (\partial_t\ex{X_i})\partial_t\ex{X_j}
        \bigg)\bigg|
        \\
        & \le\bigg|\partial_t^2\ex{X_iX_j}\bigg|  \nonumber
        \\
        &\quad + \bigg|(\partial_t^2\ex{X_i})\ex{X_j}+ \ex{X_i}\partial_t^2\ex{X_j}\bigg|         \nonumber
        \\
        & \quad + \bigg|2 (\partial_t\ex{X_i})\partial_t\ex{X_j}
        \bigg| .
        \label{e:second_derivative_size_expansion}
    \end{align}
    We will bound the terms in Eq.~\eqref{e:second_derivative_size_expansion} from bottom to top.

    The factors with only one $\partial_t$ can be handled as in Eqns.~\eqref{e:generator_definition}-\eqref{e:covariance_derivative_with_L}:
    \begin{align}
        \partial_t\ex{X_i} & = \ex{\mathcal{L}_t X_i}
        \\
        & = \ex{v_i}.
    \end{align}
    By Assumption~\ref{assumption:velocity_bound}, $|\ex{v_i}|\le C_{\ve{v}}$, so
    \begin{align}
        \bigg|2 (\partial_t\ex{X_i})\partial_t\ex{X_j}
        \bigg| \le 2C_{\ve{v}}^2.
        \label{e:second_derivative_third_term_bound}
    \end{align}

    The remaining terms involve $\partial_t^2$. Given a function $f(\ve{x})$ whose expectation is twice continuously differentiable,
    \begin{align}
        \partial_t^2\ex{f(\ve{X})} & = \langle f, \partial_t^2 p\rangle
        \\
        & = \langle f, \partial_t\mathcal{L}_t^* p\rangle
        \\
        & = \langle f, (\mathcal{L}_t^*\partial_t + [\partial_t, \mathcal{L}_t^*]) p\rangle
        \\
        & = \langle (\mathcal{L}_t^2 + [\partial_t, \mathcal{L}_t^*]^*)f, p\rangle,
        \label{e:generic_expectation_second_derivative}
    \end{align}
    where $ [\partial_t, \mathcal{L}_t^*]$ is the commutator defined by
    \begin{align}
        [\partial_t, \mathcal{L}_t^*]p & = \partial_t \mathcal{L}_t^*p - \mathcal{L}_t^* \partial_t p
        \\
        & = - \partial_i \left[(\partial_t v_i ) p\right] + \partial_i\partial_j \left[(\partial_t \D_{ij}) p\right].
        \label{e:dt_L_commutator}
    \end{align}
    Taking the adjoint,
    \begin{align}
        [\partial_t, \mathcal{L}_t^*]^* f & = \partial_t v_i \partial_i f + \partial_t \D_{ij}\partial_i\partial_j f.
    \end{align}

    For $f(\ve{x}) = x_i$, $\mathcal{L}_t^2 x_i = \mathcal{L}_t v_i$ and
    \begin{align}
        \partial_t^2 \ex{X_i} & = \ex{\mathcal{L}_t v_i + \partial_t v_i}
        \\
        & = \ex{v_k \partial_k v_i + \D_{k\ell}\partial_k\partial_\ell v_i + \partial_t v_i}.
    \end{align}
    Separating the terms and applying the Cauchy-Schwarz inequality gives
    \begin{align}
        \left|
            \partial_t^2 \ex{X_i}
        \right|
        \le
        \ex{v_k^2}^{1/2}\ex{(\partial_k v_i)^2}^{1/2} 
        + \ex{\D_{kl}^2}^{1/2} \ex{(\partial_k\partial_\ell v_i)^2}^{1/2} + |\ex{\partial_t v_i}|,
    \end{align}
    where the sums over $k$ and $\ell$ are done after the square roots.
    From Assumptions~\ref{assumption:velocity_bound}-\ref{assumption:D_bound},
    \begin{align}
        \ex{v_k^2} &\le C_{\ve{v}}^2
        \\
        \ex{(\partial_kv_i)^2} & \le C_{\ve{v}_x}^2
        \\
        \ex{(\partial_k\partial_\ell v_i)^2} & \le C_{\ve{v}_{xx}}^2
        \\
        \ex{\D_{k\ell}^2} & \le C_{\D}^2
        \\
        |\ex{\partial_t v_i}| & \le C_{\ve{v}_t}.
    \end{align}
    Hence
    \begin{align}
        \left|
            \partial_t^2 \ex{X_i}
        \right|
        \le 
        dC_{\ve{v}} C_{\ve{v}_x}
        +  d^2C_{\D} C_{\ve{v}_{xx}}
        + C_{\ve{v}_t}
    \end{align}
    and the second term in Eq.~\eqref{e:second_derivative_size_expansion} is bounded by
    \begin{align}
        \bigg|(\partial_t^2\ex{X_i})\ex{X_j}+ \ex{X_i}\partial_t^2\ex{X_j}\bigg|   \le 2 C_{\ve{x}}(dC_{\ve{v}} C_{\ve{v}_x}
        +  d^2 C_{\D} C_{\ve{v}_{xx}}
        + C_{\ve{v}_t}).
        \label{e:second_derivative_second_term_bound}
    \end{align}

    For $f(\ve{x}) = x_ix_j$, 
    \begin{align}
        \mathcal{L}_t^2 x_i x_j & = \mathcal{L}_t (x_i v_j + \D_{ij} + \swapij)
        \\
        & = (v_k \partial_k + \D_{k\ell}\partial_k\partial_\ell)(x_i v_j + \D_{ij}) + \swapij
        \\
        & = v_i v_j + x_i v_k\partial_k v_j + v_k\partial_k \D_{ij} + x_i \D_{k\ell} \partial_k\partial_\ell  v_j + \D_{i\ell} \partial_\ell  v_j + \D_{ki} \partial_k  v_j + \D_{k\ell} \partial_k\partial_\ell\D_{ij}
        \nonumber
        \\
        &\quad 
        + \swapij,
        \label{e:L2_applied_to_xi_xj}
    \end{align}
    where $\swapij$ means repeating the previous terms with indices $i$ and $j$ swapped. We also need 
    \begin{align*}
        [\partial_t, \mathcal{L}_t^*]^* x_i x_j & = x_i \partial_t v_j + \partial_t \D_{ij}
         + \swapij.
        \label{e:commutator_applied_to_xi_xj}
    \end{align*}

    Putting the pieces from Eqns.~\eqref{e:L2_applied_to_xi_xj} and \eqref{e:commutator_applied_to_xi_xj} into Eq.~\eqref{e:generic_expectation_second_derivative},
    \begin{align}
        \partial_t^2 \ex{X_iX_j} & = \ex{v_i v_j + x_i v_k\partial_k v_j + v_k\partial_k \D_{ij} + x_i \D_{k\ell} \partial_k\partial_\ell  v_j + \D_{i\ell} \partial_\ell  v_j + \D_{ki} \partial_k  v_j + \D_{k\ell} \partial_k\partial_\ell\D_{ij}}
        \nonumber
        \\
        & \quad + \ex{x_i \partial_t v_j + \partial_t \D_{ij}}\nonumber
        \\
        & \quad + \swapij.
    \end{align}
Taking the terms in order, splitting factors with the Cauchy-Schwarz inequality, and applying Assumptions~\ref{assumption:x_bound}-\ref{assumption:D_derivative_bounds},
\begin{align}
    |\ex{v_i v_j}| & \le C_{\ve{v}}^2
    \\
    |\ex{x_i v_k\partial_k v_j}| & \le \ex{x_i^4}^{1/4} \ex{v_k^4}^{1/4} \ex{(\partial_k v_j)^2}^{1/2}
    \le d C_{\ve{x}} C_{\ve{v}} C_{\ve{v}_x}
    \\
    |\ex{v_k\partial_k \D_{ij}}| & \le \ex{v_k^2}^{1/2} \ex{(\partial_k \D_{ij})^2}^{1/2}
    \le dC_{\ve{v}} C_{\D_x}
    \\
    |\ex{x_i \D_{k\ell}\partial_k\partial_\ell  v_j }| & \le 
    \ex{x_i^4}^{1/4} \ex{\D_{k\ell}^4}^{1/4} \ex{(\partial_k\partial_\ell  v_j )^2}^{1/2} 
    \le d^2 C_{\ve{x}}C_{\D} C_{\ve{v}_{xx}}
    \\
    |\ex{\D_{i\ell} \partial_\ell  v_j}| & \le \ex{\D_{i\ell}^2}^{1/2}\ex{(\partial_\ell  v_j)^2}^{1/2}\le d C_{\D} C_{\ve{v}_x}
    \\
    |\ex{\D_{ki} \partial_k  v_j}| & \le d C_{\D} C_{\ve{v}_x}
    \\
    |\ex{\D_{k\ell} \partial_k\partial_\ell\D_{ij}} | & \le \ex{\D_{k\ell}^2}^{1/2}\ex{(\partial_k\partial_\ell  \D_{ij})^2}^{1/2} \le d^2 C_{\D} C_{\D_{xx}}
    \\
    |\ex{x_i \partial_t v_j}| & \le 
        \ex{x_i^2}^{1/2} \ex{(\partial_t v_j)^2}
    ^{1/2} \le C_{\ve{x}} C_{\ve{v}_t}
    \\
    |\ex{\partial_t \D_{ij}}| & \le C_{\D_t},
\end{align}
where again the implicit sums over $k$ and $\ell$ are done last.
With an additional factor of $2$ from the $\swapij$ terms, we now have
\begin{align}
    |\partial_t^2 \ex{X_iX_j}| & \le 2 \bigg(C_{\ve{v}}^2 + d C_{\ve{x}} C_{\ve{v}} C_{\ve{v}_x} + dC_{\ve{v}} C_{\D_x} + d^2 C_{\ve{x}}C_{\D} C_{\ve{v}_{xx}}
    \nonumber
    \\
    & \qquad + 2 d C_{\D} C_{\ve{v}_x} +  d^2 C_{\D} C_{\D_{xx}} +  C_{\ve{x}} C_{\ve{v}_t} + C_{\D_t}\bigg).
    \label{e:second_derivative_first_term_bound}
\end{align}
We can now put together Eqs.~\eqref{e:second_derivative_third_term_bound}, \eqref{e:second_derivative_second_term_bound}, and \eqref{e:second_derivative_first_term_bound} in Eq.~\eqref{e:second_derivative_size_expansion} to get
\begin{align}
    \left\|\partial_t^2 \cov(\ve{X}(t))\right\|_\infty & \le 2 \bigg(C_{\ve{v}}^2 + d C_{\ve{x}} C_{\ve{v}} C_{\ve{v}_x} + dC_{\ve{v}} C_{\D_x} + d^2 C_{\ve{x}}C_{\D} C_{\ve{v}_{xx}}
     + 2 d C_{\D} C_{\ve{v}_x} +  d^2 C_{\D} C_{\D_{xx}} +  C_{\ve{x}} C_{\ve{v}_t} + C_{\D_t}\bigg)
     \nonumber
    \\
    & \quad + 2 C_{\ve{x}}(dC_{\ve{v}} C_{\ve{v}_x}
    +  d^2 C_{\D} C_{\ve{v}_{xx}}
    + C_{\ve{v}_t})\nonumber
    \\
    & \quad + 2 C_{\ve{v}}^2
    \\
    & = 2\bigg(2C_{\ve{v}}^2 + 2d C_{\ve{x}} C_{\ve{v}} C_{\ve{v}_x} + dC_{\ve{v}} C_{\D_x} + 2d^2 C_{\ve{x}}C_{\D} C_{\ve{v}_{xx}}
    + 2 d C_{\D} C_{\ve{v}_x} +  d^2 C_{\D} C_{\D_{xx}} +  2C_{\ve{x}} C_{\ve{v}_t} + C_{\D_t}\bigg).
\end{align}

Lemma~\ref{lemma:matrix_Taylors_theorem} now completes the proof.
\end{proof}

\begin{lemma}[Vector Taylor's theorem]
    Let $\ve{f}(t):\mathbb{R}\to\mathbb{R}^m$ be twice continuously differentiable in $t$. If $\|\partial_t^2 \ve{f}(t)\|_\infty \le C$ for all $t$, then 
    \begin{align}
        \left\|\ve{f}(t + \delta t) - \ve{f}(t) - \delta t \partial_t \ve{f}(t) \right\|_\infty \le \frac{C}{2} \delta t^2.
    \end{align}
    \label{lemma:matrix_Taylors_theorem}
\end{lemma}
\begin{proof}
    From Taylor's Theorem, for any index $j$ there exists $t_j \in [t, t+ \delta t]$ such that
    \begin{align}
        f_j(t + \delta t) = f_j(t) + \delta t \partial_t f_j(t) + \frac{1}{2}(\partial_t^2 f_j(t_j)) \delta t^2. 
    \end{align}
    Hence
    \begin{align}
        |f_j(t + \delta t) - f_j(t) - \delta t \partial_t f_j(t)| & = \frac{1}{2}\left|\partial_t^2 f_j(t_j)\right| \delta t^2
        \\
        & \le \frac{1}{2}\left\|\partial_t^2 \ve{f}(t_j)\right\|_\infty \delta t^2
        \\
        & \le \frac{C}{2} \delta t^2.
    \end{align}
    Then
    \begin{align}
        \left\|\ve{f}(t + \delta t) - \ve{f}(t) - \delta t \partial_t \ve{f}(t) \right\|_\infty & = \max_j |f_j(t + \delta t) - f_j(t) - \delta t \partial_t f_j(t)|
        \\
        & \le \frac{C}{2} \delta t^2.
    \end{align}
\end{proof}

\subsubsection{Velocity estimation error}
\label{appendix:velocity_error_lemma}

\begin{lemma}[Covariance combination]
    If $\|\Delta\|_\infty \le C_1$ almost surely and $\var(X_i)\le C_2^2$, then
    \begin{align}
        \|\cov(X + \Delta) -\cov(X)\|_\infty \le 2 C_1 C_2 + C_1^2.
    \end{align}
    \label{lemma:covariance_change}
\end{lemma}
\begin{proof}
    \begin{align}
        \cov(X + \Delta) & = \cov(X) + \ex{(X -\bar X)(\Delta - \bar \Delta)^\top} + \ex{(\Delta - \bar \Delta)(X - \bar X)^\top} + \cov(\Delta).
    \end{align}
    So
    \begin{align}
        |(\cov(X + \Delta) - \cov(X))_{ij}| & \le |\cov(X_i, \Delta_j)| + |\cov(\Delta_i, X_j)| + |\cov(\Delta_i, \Delta_j)|
        \\
        & \le \var(X_i)^{1/2}\var(\Delta_j)^{1/2} + \var(\Delta_i)^{1/2} \var(X_j)^{1/2} + \var(\Delta_i)^{1/2}\var(\Delta_j)^{1/2}.
    \end{align}
    Since $|\Delta_i| \le C_1$ almost surely, $\var{\Delta_i} \le C_1^2$. So
    \begin{align}
        |(\cov(X + \Delta) - \cov(X))_{ij}| & \le C_2C_1 + C_1C_2 + C_1^2.
    \end{align}
\end{proof}

\begin{lemma}[Velocity estimation error]
    If $\|\ve{\hat v}(\ve{x}, t) - \ve{v}(\ve{x}, t)\|_\infty \le C_{\ve{\hat v} - \ve{v}}$ and $\|\cov(\ve{X}(t))\|_\infty \le C_{\ve{x}}^2$, then
    \begin{align}
        \|\cov(\ve{X}(t) + \delta t \ve{\hat v}(\ve{X}(t), t)) - \cov(\ve{X}(t) + \delta t \ve{v}(\ve{X(t)}, t))\|_\infty \le 2C_{\ve{\hat v} - \ve{v}}C_{\ve{x}}\delta t + C_{\ve{\hat v} - \ve{v}}^2 \delta t^2.
        \label{e:velocity_error_lemma}
    \end{align}
    \label{lemma:velocity_error}
\end{lemma}
\begin{proof}
    Apply Lemma~\ref{lemma:covariance_change} with $C_2 = C_{\ve{x}}$ and $C_1 = C_{\ve{\hat v} - \ve{v}}\delta t$.
\end{proof}

\end{document}